\documentclass[journal, twocolumn]{IEEEtran}
\IEEEoverridecommandlockouts
\usepackage{cite}
\usepackage{amsmath,amssymb,amsfonts,bm}
\usepackage{algpseudocode}
\usepackage{algorithm,float}
\usepackage{graphicx,subfigure}
\usepackage{textcomp}
\usepackage{xcolor}
\usepackage{multirow}
\usepackage{colortbl}
\usepackage{url}
\usepackage{caption}
\usepackage{graphicx}
\usepackage{grffile}
\def\BibTeX{{\rm B\kern-.05em{\sc i\kern-.025em b}\kern-.08em
T\kern-.1667em\lower.7ex\hbox{E}\kern-.125emX}}

\definecolor{LightCyan}{rgb}{0.88,1,1}
\definecolor{LightGray}{rgb}{0.88,0.88,0.88}

\newtheorem{rem}{Remark}
\definecolor{org}{rgb}{0.9,0.4,0.1}

\begin{document}
\title{Deep Autoencoder-based  Z-Interference Channels with Perfect and Imperfect CSI} 
	
\author{Xinliang Zhang, \IEEEmembership{Member, IEEE},  Mojtaba Vaezi, 
\IEEEmembership{Senior Member, IEEE}  
\thanks{ 
	This paper has been presented in part at IEEE Wireless Communications and Networking Conference (WCNC), 2023 \cite{zhang2023DAE} and IEEE International Conference on Communications (ICC), 2023 \cite{zhang2023interference}. This work was supported by the U.S. National Science Foundation under Grants ECCS-2301778 and CNS-2239524.
	
X. Zhang and M. Vaezi are with the Department
of Electrical and Computer Engineering, Villanova University, Villanova,
PA 19085 USA (e-mail: xzhang4@villanova.edu; mvaezi@villanova.edu).
} 
}

\maketitle

\begin{abstract}
A deep autoencoder (DAE)-based structure for end-to-end communication over the  two-user Z-interference channel (ZIC) with finite-alphabet inputs is designed in this paper. The proposed structure jointly optimizes the two encoder/decoder pairs and  generates interference-aware constellations that dynamically adapt their shape based on interference intensity to minimize the bit error rate (BER). An in-phase/quadrature-phase (I/Q) power allocation layer is introduced in the DAE to guarantee an average power 
constraint and enable the architecture to generate constellations with nonuniform shapes. This  brings further gain compared to standard uniform constellations such as quadrature amplitude modulation. The proposed structure is then extended to work with imperfect channel state information (CSI). The CSI imperfection due to both the estimation and quantization errors are examined.    
The performance of the  DAE-ZIC is compared with two baseline methods, i.e., standard and rotated 
constellations. The proposed structure significantly enhances the performance of the ZIC both for the perfect and imperfect CSI. Simulation results show that the improvement is achieved
in all interference regimes (weak, moderate, and strong) and consistently increases with the signal-to-noise ratio (SNR). For instance, more than an order of magnitude BER reduction is obtained with respect to the most competitive conventional method at weak interference when $\rm SNR>15 dB$
and two bits per symbol are transmitted.
The improvements reach about two orders of magnitude when quantization error exists, indicating that the DAE-ZIC is more robust to the interference compared to the conventional methods.  
\end{abstract}


\section{Introduction}

Interference is a central issue in today's \textit{multi-cell} networks.
The information-theoretic model for a multi-cell network is the \textit{interference channel} (IC). 
There have been many efforts to find the capacity of the IC either 
with 
the same generality and accuracy used by Shannon for point-to-point systems 
\cite{carleial1975case,sato1981capacity,han1981new,sason2004achievable} or 
by seeking approximate solutions with a guaranteed gap to optimality at any signal-to-noise
ratio (SNR) \cite{etkin2008gaussian}.
However, the capacity region of the two-user IC is only  known for strong  
interference   where decoding and canceling the 
interference is optimal \cite{carleial1975case}. 
Also, at very 
weak interference,  sum-capacity is achievable by treating interference as noise 
\cite{motahari2009capacity, shang2009new, annapureddy2009gaussian}, whereas, in general, decoding part of the interference  and treating the 
remaining as noise is the best achievable scheme to date \cite{han1981new}.

The aforementioned Shannon-theoretic works are based on  Gaussian inputs.  Despite being  
theoretically optimal, Gaussian alphabets are continuous and unbounded, and thus, are  rarely 
applied in real-world communication. 
In practice, signals are  generated using finite alphabet sets, 
such as phase-shift keying (PSK) 
and  quadrature amplitude modulations (QAM). 
The performance gap between the finite alphabet input  and the Gaussian input design is non-negligible \cite{wu2013linear}. However, conventional finite-alphabet approaches are based on predefined uniform constellations like QAM. 
These constellations are defined for point-to-point systems 
\cite{foschini1974optimization,forney1984efficient,goldsmith1997variable,barsoum2007constellation} and their constellation shaping is oblivious to interference. Such an inability to respond to interference is an obstacle to improving the bit-error rate and spectral efficiency of today’s interference-limited communication systems.



In this paper, we consider the two-user single-input 
single-output (SISO) one-sided IC,   also known as the Z-interference channel (ZIC).
With Gaussian signaling, the capacity region of this channel is known only in  the strong and very strong interference regimes  
\cite{sason2004achievable}.   
However, Gaussian signaling is unsuitable for practical applications.  
%
%
Previous work has studied ZIC with finite alphabet sets in specific regimes and predefined uniform constellations.   In \cite{knabe2010achievable}, it is shown that rotating one input constellation (alphabet) can improve the sum-rate of the two-user IC in strong/very strong interference 
regimes. 
Later,  an exhaustive search  for finding the optimal rotation of the signal 
constellation  
was presented in \cite{ganesan2012two}.   
In addition, a signaling design is proposed in \cite{ho2012improper} which applies a rotation to the channel which resembles rotating the input. 
The main focus of the above papers is to maximize the achievable rates, and they do not study bit-error rate (BER) performance. However, BER is a key metric and interference can severely reduce the BER by distorting the received constellation, when uniform constellations like QAM are employed.


Recent research has proved end-to-end learning as a promising approach for encoding, decoding, and signal representation to reduce BER \cite{o2017introduction}.
Particularly, deep autoencoder (DAE) is a popular architecture for implementing end-to-end learning. It consists of an encoder that transforms input data into a low-dimensional representation to find its \textit{structure} and a decoder that reconstructs the original input from this representation. 
{DAE-based} end-to-end  communication is an emerging approach to  finite-alphabet communication in which BER is the main performance measure and constellation design is inherent.  DAE-based communication is introduces both for single- and multi-user systems by various groups \cite{o2017introduction,o2017deep,nartasilpa2018communications,ye2020deep,song2020benchmarking,zhang2021svd}. 	 These studies indicate that the DAE surpasses current solutions and enhances performance beyond conventional methods \cite{zhang2021svd}.

	Specifically, utilizing two DAEs for the transmitter/receiver pairs enables effective signal separation/decoding of original data even in the presence of interference, thereby paving the way for enhancing communication performance over the IC, as investigated in prior studies \cite{o2017introduction, erpek2018learning, wu2020deep}. However, these studies have limitations as they focus solely on symmetric interference scenarios and compare their results against simple baselines like quadrature phase shift keying (QPSK), despite the fact that QPSK performs much worse than rotated QPSK in the context of  the IC \cite{knabe2010achievable,ganesan2012two,ho2012improper}.
Further, the DAE designs in \cite{o2017introduction, erpek2018learning, wu2020deep} produce symbols with fixed power levels and lack the ability to generalize to QAM-like constellations. Thus,   they do not efficiently use the in-phase and quadrature-phase (I/Q) plane. Additionally, these designs assume perfect knowledge of channel state information (CSI), and the transition from perfect CSI to imperfect CSI remains unexplored.

\subsection{Motivation and Contribution}
The above limitations has motivated us to investigate DAEs potential
	for the long-lasting problem of interference in more practical settings. 
We shed light on DAE-based communication over
asymmetric interference with both perfect and imperfect CSI.
Specifically, we design and train novel DAE-based architectures for the ZIC with 
finite-alphabet inputs.  In the ZIC, two DAEs should be 
considered in two  transmitter-receiver pairs. 
The two DAE pairs cooperate to avoid interference
and adapt their constellation 
to the interference intensity. Our work is motivated by the following question: 
\textit{Can we design interference-aware constellations using DAEs}?  
Will the gains remain/vanish if CSI is not perfect? 
We answer these questions by developing new structures and explaining  
how the designed \textit{nonuniform constellations}  lend themselves to interference mitigation in different regimes, and thus 
improving the BER.  

%
The ZIC  is characterized by local, short-range interference \cite{shamai2011rate}, where far-away users are not affected by interference. This simple channel model is a fundamental building block for more complex interference networks and its understanding is crucial in the field of interference research.
	As such, it has been widely studied in the literature of interference \cite{sason2004achievable,kramer2004review,vaezi2016simplified}, as it provides insights into the limits interference-limited scenarios. By understanding the behavior of interference in  the ZIC, we can develop more effective strategies to mitigate interference in more complex networks.  

The main contributions of the paper are as follows: 
\begin{itemize}
\item We design a DAE-based transmission structure for the ZIC and  demonstrate its effectiveness across  weak, moderate, and strong  interference levels.
We propose incorporating an average power constraint normalization layer that enables \textit{nonuniform} constellations, resulting in more efficient utilization of the I/Q plane.   
The designed constellations are adaptive to 
the interference intensity, and morph in a way that the `receivers' see distinguishable symbols, thereby improving BER even in the presences of interference.  
We also conduct a neural network ablation study to demonstrate the impact and necessity of each design element in our proposed model. This analysis provides valuable insights into the significance of each component and also serves to validate the overall architectural effectiveness.  

\item We extend the proposed structure to the finite-alphabet ZIC with imperfect CSI, where we consider both estimation and quantization errors. Particularly, CSI estimation errors pose confuses  for DAE training and testing performance. Meanwhile, quantization errors, arising from limited feedback capacity, introduce undesirable rotations to the constellations.  In order to develop a DAE that is robust against these errors, we introduce an equivalent system model that reduces the CSI parameters and ultimately lowers the BER.
 \item Our design directly accepts the transmission bits as its input rather than converting them to 
 symbols and using one-hot vectors for the DAE input. This has two advantages. First, our design can 
 directly minimize the BER and we do not need to worry about optimal bit-to-symbol mapping. Second, 
 it reduces the complexity as to transmit $b$ bits, the input and output layers require only $b$ neurons 
 whereas the one-hot vector method needs $2^b$ neurons.      
\end{itemize}

It worth mentioning that, for benchmarking  purposes, we use rotated uniform constellations, which have been proven to be more competitive than their unrotated counterparts.  Our  proposed DAE-ZIC shows significantly better BER  
performance for all interference regimes (weak, 
moderate, and strong interference). 
For perfect CSI, when averaged over SNRs from 0 to 20dB, 44\% reduction in BER is achieved.
At certain SNRs and interference regimes, the improvement is over an order of magnitude.   
For imperfect CSI, the overall reduction is about 40\%. The gap between the DAE-ZIC and conventional methods is even larger when quantization error is applied.    

\subsection{Organization}
The remainder of this paper is organized as follows. We elaborate on the 
ZIC system model in Section~\ref{sec_sys}. 
The  DAE  design and the training approach are introduced in Section~\ref{sec_dae}. The system model with an imperfect CSI is derived in Section~\ref{sec_impCH}. The modifications for the DAE with an imperfect CSI is introduced in Section~\ref{sec_DAE_impCh}. 
Numerical results are presented in  Section~\ref{sec_simulation}, and 
the paper is concluded in Section~\ref{sec_con}.

\textit{Notation:}
$(\cdot)^{T}$ 
 denotes transpose, 
 $\mathbb{E}\{\cdot\}$ denotes expectation, and ${\rm{diag}}(\lambda_1, 
\dots, \lambda_n)$ represents the diagonal matrix with elements  
$\lambda_1, 
\dots, \lambda_n$. 
$\mathcal{N}(\mu,\sigma^2)$ and $\mathcal{CN}(\mu,\sigma^2)$ are real and complex Gaussian distributions
where $\mu$ and $\sigma^2$ are the mean and variance.  $|\cdot|$ is the amplitude  of a complex number. 

\section{System Model of  ZIC with Perfect CSI}\label{sec_sys}
\begin{figure}[tb] 
\centering
\includegraphics[scale=.64]{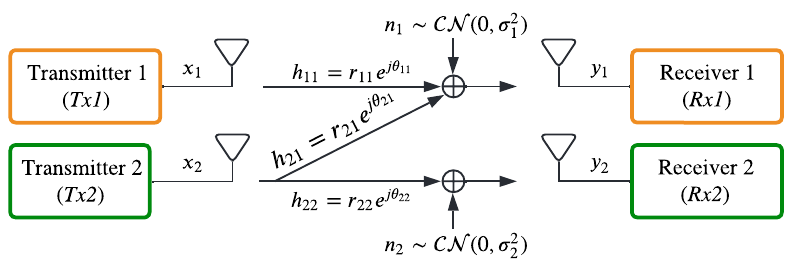}
\caption{System model of the ZIC.}
\label{fig_sys1} 
\end{figure}

Figure~\ref{fig_sys1} shows the system model of a two-user complex SISO ZIC with perfect CSI.  The two 
transmitter-receiver pairs  wish to 
reliably transmit their messages while
the transmission of the first pair  interferes with the
transmission of the second. 
The four nodes are named 
\textit{Tx1}, \textit{Tx2}, \textit{Rx1}, and \textit{Rx2}, as shown in 
Fig.~\ref{fig_sys1}.  $h_{ij}$ refers to the channel gain from the $i$th transmitter to 
the $j$th receiver and $i,j\in\{1,2\}$. For ZIC,  $h_{21}=0$.  The received signals at the 
receivers can be written as 
\begin{subequations}\label{eq_ZIC_sys}
\begin{align}
&y_1 = h_{11}x_1 + h_{21}x_2 + n_1,\\
&y_2 = h_{22}x_2  + n_2,
\end{align}
\end{subequations}
in which $x_1$ and $x_2$ denote the transmitted symbols of \textit{Tx1} and \textit{Tx2}. 
The transmitted signals are complex-valued with finite-alphabets 
and variances $\mathbb{E}\{|x_1|^2\}=P_1$ and $\mathbb{E}\{|x_2|^2\}=P_2$ in which $P_1$ and 
$P_2$ are the power budgets of the two transmitters. The channel coefficients are complex  random variables 
\begin{align}
h_{ij}\triangleq r_{ij}e^{j\theta_{ij}}\sim\mathcal{CN}(\mu_h, \sigma_h^2),\label{eq_perfectCH}
\end{align}
where $\mu_h$ and $\sigma_h^2$ are the mean and variance of the channel distribution, and  $r_{ij}$ and $\theta_{ij}$ represent the magnitude and phase of $h_{ij}$. Also, 
 $n_{1}$ and  $n_{2}$ are the complex-valued  independent 
and identically distributed (i.i.d.) additive white Gaussian noise with zero means and variances 
$\sigma_1^2$ 
and $\sigma_2^2$. Without loss of generality, we assume 
the noise powers at the two receivers' sides are the same, i.e., $\sigma_1^2=\sigma_2^2=\sigma^2$ 
\cite{kramer2004review}.

\subsection{The Equivalent System Model of the ZIC} \label{sec_sysA}
\begin{figure}[tb] 
	\centering
	\subfigure[Original system model]
	{\includegraphics[scale=.64]{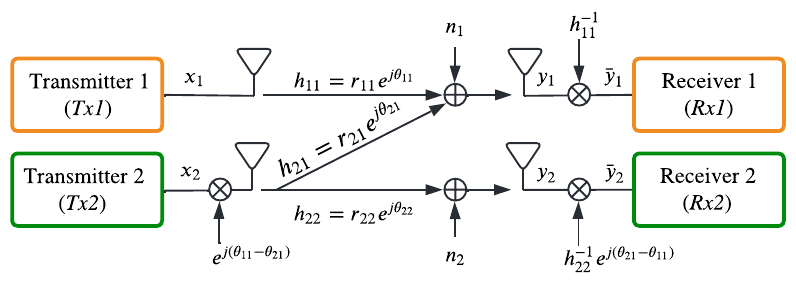}}
	\subfigure[Equivalent system model]
	{\includegraphics[scale=.64]{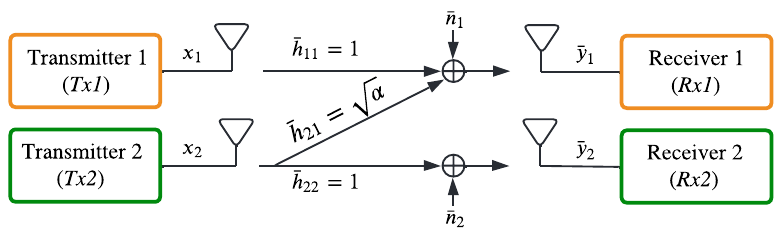}}
	\caption{The equivalent system model of the ZIC.}
	\label{fig_sys2} 
\end{figure}

It is known that, without loss of generality, the channel 
gains of the direct transmission links can be modeled as one, shown in Fig.~\ref{fig_sys2}(b), \cite{vaezi2016simplified, 	lameiro2016rate, kramer2004review}. The 
interference gain is also real-valued  for both real- 
and complex-valued systems.   When CSI is 
available and if we apply pre- and post-processing  illustrated in Fig.~\ref{fig_sys2}(a),  such a 
system model in Fig.~\ref{fig_sys1}  is equivalent to that of Fig.~\ref{fig_sys2}(b).\footnote{We describe this process here as we will need later in Section~\ref{sec_impCH} where CSI is not perfect. } 
The \textit{Tx2} applies $e^{j(\theta_{11}-\theta_{21})}$ to cancel the phase of $h_{21}$ and align its phase with that of $h_{11}$.  The
\textit{Rx1} and \textit{Rx2} applies $h_{11}^{-1}$ and 
{$h_{22}^{-1}e^{j(\theta_{21}-\theta_{11})}$} to normalize 
the channel gain to one. Then, the received  post-processed signals are
\begin{subequations}
\begin{align}\label{eq_ybar}
&\bar{y}_1=h_{11}^{-1} y_{1}=x_1+r_{21}r_{11}^{-1}x_2+n_1h_{11}^{-1},\\
&\bar{y}_2=h_{22}^{-1}e^{j\theta_{21}} y_{2}=x_2+e^{j\theta_{21}}{n_1}h_{22}^{-1}.
\end{align}
\end{subequations}
By defining 
\begin{subequations}
\begin{align}
&\sqrt{\alpha}\triangleq r_{21}r_{11}^{-1},\\
&\bar{n}_1\triangleq n_1h_{11}^{-1},\;\textmd{and}, \;
\bar{n}_2\triangleq e^{j\theta_{21}}{n_1}h_{22}^{-1},\label{eq_eqNoise}
\end{align}
\end{subequations}
we have the system model in Fig.~\ref{fig_sys2}(b) as
\begin{subequations} \label{eq_sysModel}
\begin{align} 
&\bar{y}_1=\bar{h}_{11}x_1+\sqrt{\alpha}x_2+\bar{n}_1,\\
&\bar{y}_2=\bar{h}_{22}x_2+\bar{n}_2.
\end{align}
\end{subequations}
where $\bar{h}_{11}=\bar{h}_{22}=1$ and $\bar{h}_{21}=\sqrt{\alpha}$ are the equivalent channel gains.

Thus, the two system models in Fig.~\ref{fig_sys1} and Fig.~\ref{fig_sys2} are 
equivalent. {Hence, we follow the existing studies and use the system model in 
Fig.~\ref{fig_sys2}(b),}  
and consider a fixed $\sqrt{\alpha}$ at 
each time. 
It is worth mentioning that both  actual channel gains and noise {($h_{ij}$ and  $n_i$, $i,j\in\{1,2\}$)} are 
Gaussian. In this paper, we assume a slow fading scenario with
\begin{align}
\bar{n}_i\sim\mathcal{CN}(0, \sigma_{i}^2r_{ii}^{-2}). \label{eq_perfCH_noise}
\end{align}

\section{Deep Autoencoder with Perfect CSI}\label{sec_dae}
\begin{figure*}[htbp] 
	\centering
	\includegraphics[scale=.544, trim=12 10 14 20, 
	clip]{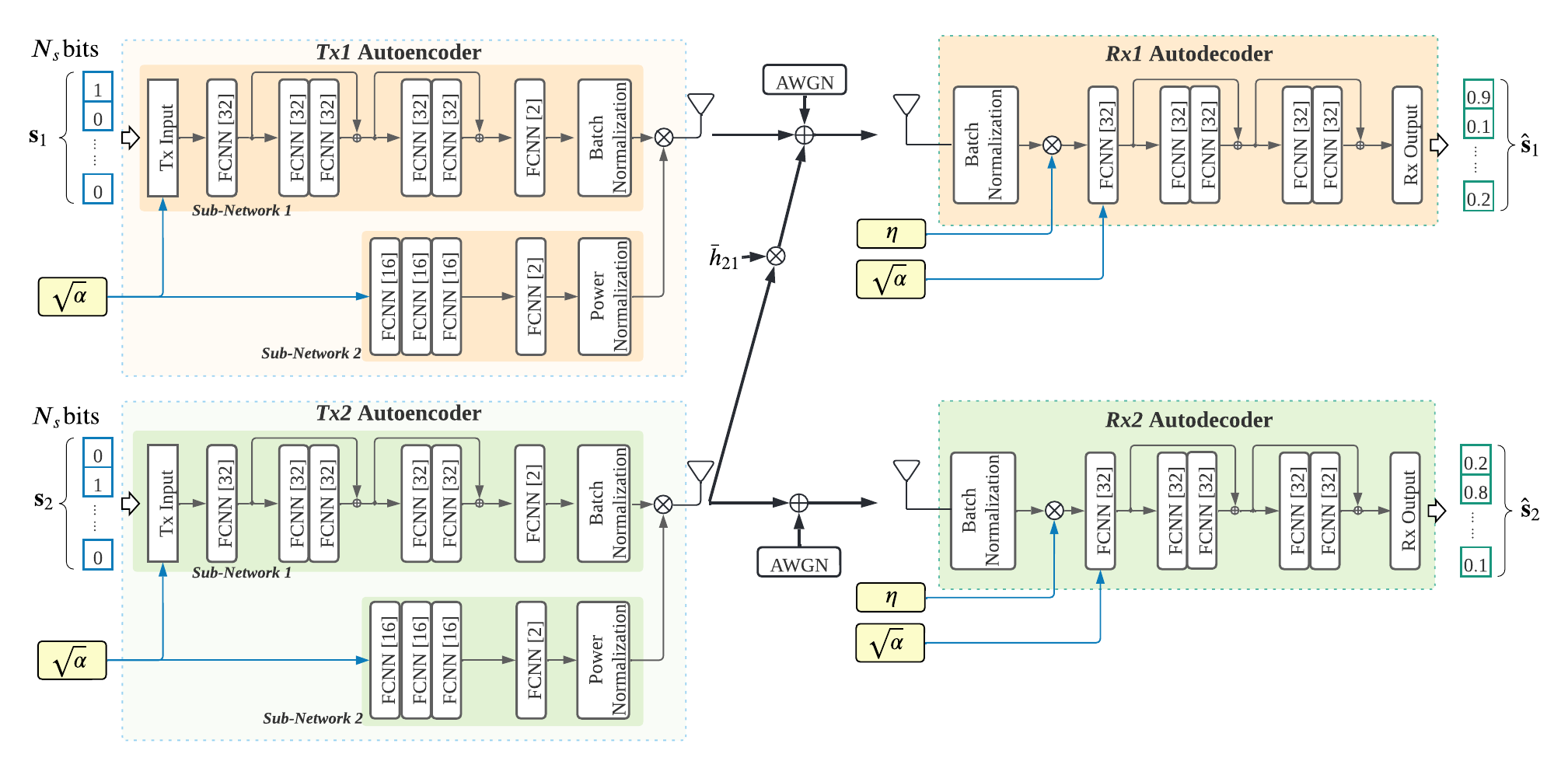}
	\caption{The architecture of the two-user  DAE-ZIC implemented by two pairs of 
		deep autoencoders. Each transmitter of the ZIC contains two sub-networks. 
		\textit{Sub-network~1}
		mainly 
		generates the constellation and \textit{sub-network 2} is used to implement the
		average power constraint. The receivers decode their bits from the received signal. This 
		architecture is based on the system model in  Fig.~\ref{fig_sys2}(b). $\eta$ 
		is 	a power control parameter defined in 
		\eqref{eq_eta}.}
	\label{fig_DAE_net} 
\end{figure*}

Existing studies \cite{knabe2010achievable, 
	ganesan2012two, deshpande2009constellation} use standard QAM constellations at each transmitter.
	Such constellations are fixed and are not adjustable according to the  interference intensity. 
	To further improve the transmission performance, we propose a DAE-based transmission for 
	the two-user ZIC, named DAE-ZIC. 
The architecture  is shown in Fig.~\ref{fig_DAE_net}.  

\subsection{The Architecture of DAE-ZIC}
The DAE-ZIC consists of two pairs of DAEs. Each pair performs an end-to-end 
transmission,  which includes input bits, 
autoencoder at the transmitter, 
channel and noise layers, autoencoder at the receiver, and final output bits. 

\subsubsection{Network Input}
Each transmitter sends $N_s$ bits to the corresponding receiver.  The interference channel 
coefficient $\sqrt{\alpha}$  is known at the transmitter and receiver and is appended to the input 
bit 
vector. Then, both the transmitters and receivers have  the 
knowledge of the CSI. The two transmitters are expected to jointly design their 
constellations and the receivers will decode correspondingly.

\subsubsection{Transmitter DAE}
As shown in Fig.~\ref{fig_DAE_net}, the DAE of the transmitter contains two 
sub-networks: \textit{Sub-network~1} and \textit{Sub-network~2}. 
{\textit{Sub-network~1} converts the input bit-vector to symbols with unit power in the I/Q components.  \textit{Sub-network~2} performs power allocation, which controls the 
power of the I/Q components. The two sub-networks combines in parallel to yield a nonuniform constellation.} The batch normalization in \textit{sub-network 1} {together with} 
\textit{sub-network~2} realize the average power constraint at the transmit antenna. 
Having an average power constraint is necessary especially for SISO systems. In this 
way, the  I/Q plane is used efficiently, like QAM.  Otherwise, the DAE can only produce 
constant-power constellations, like PSK, where constellation points are on a circle which is not  efficient in terms of BER.

The components of \textit{sub-network 1} are fully connected layers 
(FCNN), residual connections, or shortcuts, to alleviate the vanishing gradient effect, and  the output batch 
normalization layer.
 The activation function of the FCNN 
layers is \textit{tanh}, except for the last 
layer which has two hidden nodes and no activation function.  

Let the batch size be $N_B$. The output of the last FCNN is  
\begin{align}
	\mathbf{X}_{\textmd{fcnn}}\triangleq [\mathbf{x}_{\textmd{fcnn}}^{\textmd{I}}, \;
	\mathbf{x}_{\textmd{fcnn}}^{\textmd{Q}}],
\end{align}
where $\mathbf{x}_{\textmd{fcnn}}^{\textmd{I}}$ and  
$\mathbf{x}_{\textmd{fcnn}}^{\textmd{Q}}\in\mathbb{R}^{N_B\times 1}$ are the outputs of the 
two hidden nodes and  represent I/Q of the complex-valued signal. 
Since the FCNN has unbounded outputs, it cannot guarantee a power constraint at the 
transmitter. We propose the transmitter design shown in Fig.~\ref{fig_DAE_net} to achieve an average power constraint at each 
antenna. First, we  use batch normalization in \textit{sub-network 1}  to unify  the 
average power of I/Q independently.  
 The batch normalization layer linearly normalizes  
${\mathbf{x}}_{\textmd{fcnn}}^{\textmd{I}}$ and ${\mathbf{x}}_{\textmd{fcnn}}^{\textmd{Q}}$, in which the
normalized vectors  
$\mathbf{x}_B^{\textmd{I}}$ and $\mathbf{x}_B^{\textmd{Q}}$ are given by
\begin{align}
\mathbf{x}_B^{\textmd{I}}\triangleq\beta^{\textmd{I}}\cdot{\mathbf{x}}_{\textmd{fcnn}}^{\textmd{I}}, \; \; 
\mathbf{x}_B^{\textmd{Q}}\triangleq\beta^{\textmd{Q}}\cdot{\mathbf{x}}_{\textmd{fcnn}}^{\textmd{Q}},
\end{align}
where $\bm{\beta}\triangleq[\beta^{\textmd{I}},\;\beta^{\textmd{Q}}]^T$ contains two factors for 
normalization. 

{In \textit{sub-network 2}, the power allocation of the I/Q components is determined by the FCNN layers, which take the input value $\sqrt\alpha$ into account. The FCNN layers calculates the powers of I/Q components, and the power normalization block limits the total power to $P_t$, thus achieving the intended power control.} The powers of $\mathbf{x}_B^{\textmd{I}}$ and $\mathbf{x}_B^{\textmd{Q}}$  (i.e., each batch of I/Q signals) 
are multiplied by $\gamma^{\textmd{I}}$ and  $\gamma^{\textmd{Q}}$, which are the outputs of \textit{Sub-network~2}.   Defining the input and output of the power normalization layer as ${\bm\gamma}_0\triangleq[\gamma^{\textmd{I}}_0,  
\;\gamma^{\textmd{Q}}_0]^T\in\mathbb{R}^{2\times1}$ and 
${\bm\gamma}\triangleq[\gamma^{\textmd{I}}, 
\;\gamma^{\textmd{Q}}]^T\in\mathbb{R}^{2\times1}$ respectively,
the power normalization layer normalizes  its input and scales its power, i.e.,  
	${\bm\gamma}=\sqrt{P_t}\frac{{\bm\gamma}_0}{|{\bm\gamma}_0|}$. 
Thus,
${\bm\gamma}^T{\bm\gamma}=P_t$. 
Such an operation can be done via the 
\textit{Lambda layer} in \textsc{Keras} \cite{chollet2017kerasR}. Finally, the outputs of the batch 
normalization and power normalization are multiplied together, 
i.e.,
\begin{align}
\mathbf{x}_\textmd{out}^{\textmd{I}} \triangleq 
\gamma^{\textmd{I}}\cdot\mathbf{x}_B^{\textmd{I}},\; \; 
\mathbf{x}_\textmd{out}^{\textmd{Q}} \triangleq 
\gamma^{\textmd{Q}}\cdot\mathbf{x}_B^{\textmd{Q}}.
\end{align}
The powers of $\mathbf{x}_\textmd{out}^{\textmd{I}}$ and 
$\mathbf{x}_\textmd{out}^{\textmd{Q}}$  are 
$\gamma^{\textmd{I}}$ and  $\gamma^{\textmd{Q}}$, respectively. 
In short, {batch normalization is applied to I and Q components separately and along the time axis (batch by batch) while power normalization scales the power of I/Q components at each time to reach the 
average power constraint.} That is, the 
two normalization operations are implemented in different dimensions.  The final output of the transmitter is 
\begin{align}
\mathbf{X}&\triangleq[\mathbf{x}_\textmd{out}^{\textmd{I}} , 
\;\mathbf{x}_\textmd{out}^{\textmd{Q}} 
]\notag\\
&=[\mathbf{x}_B^{\textmd{I}},\; \mathbf{x}_B^{\textmd{Q}}] \cdot 
\textmd{diag}({\bm\gamma})\notag\\
&=[\mathbf{x}_{\textmd{fcnn}}^{\textmd{I}}, \;
\mathbf{x}_{\textmd{fcnn}}^{\textmd{Q}}]\cdot\textmd{diag}({\bm\beta})\cdot\textmd{diag}({\bm\gamma}).
\end{align}
where $\mathbf{x}_{\textmd{fcnn}}^{\textmd{I}}$ and $\mathbf{x}_{\textmd{fcnn}}^{\textmd{Q}}$ 
are the  
output of the FCNN layer in \textit{sub-network 1} and represent the preliminary I/Q signals, 
$\bm\beta$ 
normalizes the power of each batch, and $\bm\gamma$ controls the power of the I/Q signals such that average power is $P_t$. 


\begin{figure}[tb] 
\centering
\includegraphics[scale=.5, trim=12 14 15 12, 
clip]{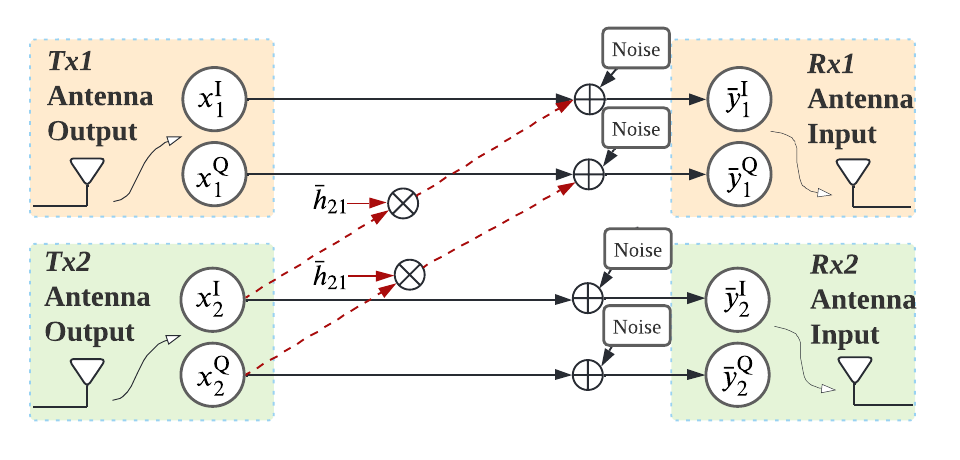}
\caption{Form a SISO complex-valued channel by a 2x2 MIMO real-valued channel.}
\label{fig_2x2realCh} 
\end{figure}
\subsubsection{Channel Implementation} 
The channel is formed following \eqref{eq_ZIC_sys}. The complex-valued 
SISO channel is achieved by real values intuitively shown in Fig.~\ref{fig_2x2realCh}. At 
\textit{Rx1},  the 
received  signal is $\bar{y}_1=\bar{y}_{1}^{\textmd{I}}+j 
\bar{y}_{1}^{\textmd{Q}}$,
\begin{align}\label{eq_DAE_ch1}
\left[\begin{matrix}
\bar{y}_{1}^{\textmd{I}}\\ \bar{y}_{1}^{\textmd{Q}}
\end{matrix}\right]=
\left[\begin{matrix}
x_{1}^{\textmd{I}}\\ x_{1}^{\textmd{Q}}
\end{matrix}\right]+\bar{h}_{21}\cdot
\left[\begin{matrix}
x_{2}^{\textmd{I}}\\ x_{2}^{\textmd{Q}}
\end{matrix}\right]+
\left[\begin{matrix}
\bar{n}_{1}^{\textmd{I}}\\ \bar{n}_{1}^{\textmd{Q}}
\end{matrix}\right],
\end{align}
where $\bar{y}_{1}^{\textmd{I}}$ and $ \bar{y}_{1}^{\textmd{Q}}$ are the I/Q components of the 
received 
signal, $x_{i}^{\textmd{I}}$ 
and $ 
x_{i}^{\textmd{Q}}$ 
are 
the I/Q components of the $i$the transmitter, and $\bar{n}_{i}^{\textmd{I}}, 
\bar{n}_{i}^{\textmd{Q}}\sim\mathcal{N}(0,\frac{1}{2}\sigma_N^2r_{ii}^{-2})$ are
the I/Q components of the complex-valued. Similarly, the 
received signal 
for 
\textit{Rx2} is $\bar{y}_2=\bar{y}_{2}^{\textmd{I}}+j \bar{y}_{2}^{\textmd{Q}}$,
\begin{align}\label{eq_DAE_ch2}
\left[\begin{matrix}
\bar{y}_{2}^{\textmd{I}}\\ \bar{y}_{2}^{\textmd{Q}}
\end{matrix}\right]=
\left[\begin{matrix}
x_{2}^{\textmd{I}}\\ x_{2}^{\textmd{Q}}
\end{matrix}\right]+
\left[\begin{matrix}
\bar{n}_{2}^{\textmd{I}}\\ \bar{n}_{2}^{\textmd{Q}}
\end{matrix}\right].
\end{align}
The additive Gaussian white noise (AWGN) is implemented by a \textit{Gaussian noise layer}  in 
\textsc{Keras}. The noise power is set according to SNR in the training and testing stages.

\subsubsection{Receiver DAE}
The received signals are $\bar{y}_1$ and $\bar{y}_2$. To ensure the receiver 
networks have a finite input range, we use \textit{batch normalization layers} in \textsc{Keras} 
unifying the power of the received signals, i.e.,    
\begin{align}\label{eq_rx_BN}
{y}_{\textmd{B},i} = \xi\cdot\bar{y}_i,\;\; E\{|{y}_{\textmd{B},i}|^2\} = 1,\;\; \forall i\in\{1,2\},
\end{align}
where $\xi$ is a coefficient to reach the  unit power. The process details and settings are 
the same as those in the transmitter.

We further define the \textit{desired signal} for \textit{Rx1} as  	 
\begin{align}
x_{\textmd{D},1} \triangleq x_1+\sqrt{\alpha}x_2.
\end{align}
$x_{\textmd{D},1}$ contains the true desired signal $x_1$ and the interference 
$\sqrt{\alpha}x_2$. The goal of the receiver is to decode $x_1$ for an arbitrary  $x_2$ in its 
constellation. The \textit{desired signal} of \textit{Rx2} is  
$x_{\textmd{D},2} \triangleq x_2$.
However, the normalization of the received signal \eqref{eq_rx_BN} causes the power of the \textit{desired signal} to vary with the SNR. Hence, the autoencoder should adjust the decoding 
boundary according to the SNR, which is an extra burden. 
So, we turn to normalize the desired signal using linear 
factor, $\eta$,  multiplied 
on the batch normalization output, i.e.,
\begin{align}\label{eq_eta}
y_{\textmd{D},i} = \eta \cdot {y}_{\textmd{B},i},\;
\eta\triangleq\sqrt{1+\frac{P_{\textmd{D,i}}}{\sigma_N^2}},\quad \forall i\in\{1,2\},
\end{align}
where $P_{\textmd{D},i}$ is the power of the \textit{desired  signal} $x_{\textmd{D},i}$ 
and 
$\sigma_N^2$ is the noise power. The batch normalization normalizes the 
\textit{desired  signals} using pre-processing $\eta$.  
Then, the normalized signal, $y_{\textmd{D},i}$, together with the feature of the ZIC, 
$h_{21}=\sqrt{\alpha}$, are sent to the rest of the FCNN layers.  
The final output of the DAE is an estimation of the transmitted bit-vectors, $\hat{\mathbf{s}}_1$ 
and $\hat{\mathbf{s}}_2$, as shown in Fig.~\ref{fig_DAE_net}.

The activation function of the output layer is sigmoid function, i.e., $f(s)=\frac{1}{1+e^{-s}}$. 
Thus, assuming the input of the final layer, marked as \textit{Rx Output} in Fig.~\ref{fig_DAE_net}, of $i$th receiver is $\mathbf{y}_{F,i}$, the output equations   can be written as $
\hat{\mathbf{s}}_{i} = f(\mathbf{y}_{F,i})$. The sigmoid function is commonly used because it outputs a value between 0 and 1, which can be interpreted as a probability of the input belonging to the positive class. This function is also differentiable, which is important for backpropagation during training of neural networks.

\subsubsection{Loss Function}
In our DAE-ZIC, each receiver has its own estimation of the transmitted bits. Then, the  overall 
loss 
function of the DAE-ZIC is 
$\mathcal{L}=\mathcal{L}_1+\mathcal{L}_2$,
where $\mathcal{L}_1$ and $\mathcal{L}_2$ are the loss  at \textit{Rx1} and \textit{Rx2}. 
Each output vector in the two receivers represents binary messages, then the network can be trained using binary cross-entropy loss: 
\begin{align}
\mathcal{L}_i=\frac{1}{N_B}\sum_{n=1}^{N_B}    (\mathbf{s}_{i,n})^T\log\hat{\mathbf{s}}_{i,n} +
(1-\mathbf{s}_{i,n})^T\log(1-\hat{\mathbf{s}}_{i,n}),
\end{align}
where $i\in\{1,2\}$ distinguishes  the users,  $N_B$ is the batch size, $\mathbf{s}_{i,n}$ is the 
$n$th  
input bit-vector in the batch, and $\hat{\mathbf{s}}_{i,n}$ is the corresponding  output. The 
loss function treats each element of the DAE output as a 0/1 classification task. The binary cross-entropy loss function is commonly used for multi-label classification problems, where each example can have multiple binary labels. The loss function measures the difference between the predicted probability of each label being present in the example and the true probability of the label being present.  
Finally, the loss is the summation of the loss of $N_s$ tasks, where $N_s$ is the number of bits in the transmission.
In the training process, the backpropagation algorithm passes $\mathcal{L}_1$ to \textit{Rx1} 
and it will further go to \textit{Tx1} and \textit{Tx2}, whereas   $\mathcal{L}_2$  affects
\textit{Rx2} and \textit{Tx2}.

\subsection{Training Procedure of the DAE-ZIC}\label{sec_train}
Due to the difficulty of training a single network across all values of the interference gain $\alpha$, distinct instances of DAEs are employed for a few different ranges of   $\alpha$. For each training session, a value for $N_s$ is selected and the desired range for $\alpha \in [\alpha_{\min}, \alpha_{\max}]$ is specified. During the training process, all four sub-networks are trained simultaneously. The DAE is trained iteratively using random values of $\alpha$ within this interval. For each $\alpha$, the DAE undergoes training for $E_p$ epochs with a mini-batch size of $N_B$ and a constant learning rate of $l_r$. After training the DAE for $N_d$ different values of $\alpha$, the learning rate is reduced to $d_r l_r$. A comprehensive description of the training procedure, including the simultaneous training of all sub-networks to adapt to the ZIC interference patterns, can be found in Algorithm~\ref{alg_Train}. 
\begin{algorithm}[htbp]
	\caption{Training Procedure  for the DAE-ZIC}\label{alg_Train}
	\begin{algorithmic}[1]
		\State Fix $N_s$, $\alpha_{\min}$, and $\alpha_{\max}$.
		\State Set $P_t=1$W and SNR$=10$dB.
		\State Set $N_\alpha=30,000$, the number of channels.
		\State Set $E_p=10$, the number of the epochs.
		\State Set $N_B=10^4$, the batch size.
		\State Set $l_r=10^{-2}$, the initial learning rate, which will  drop to $d_rl_r=0.95l_r$ after 
		every $N_d=200$ trained channels.
		\State Initialize the DAE-ZIC network.
		\For {index $i_\alpha$ from $1$ to $N_\alpha$}
		\State Uniformly and randomly select one $\alpha\in[\alpha_{\min}, \alpha_{\max}]$.
		\State {Randomly generate $h_{11}$ and $h_{22}$ using \eqref{eq_perfectCH}.
		\State Normalize the channel using \eqref{eq_sysModel}.
		\State Update $\bar{h}_{21}=\sqrt{\alpha}$ in the DAE-ZIC in 
		\eqref{eq_DAE_ch1}.
		\State Set the variance of the noise layer according to \eqref{eq_perfCH_noise}.}
		\For {index $i_e$ from $1$ to $E_p$}
		\State {Randomly  generate $N_B$ bit vectors.}
		\State Update the weights of the DAE-ZIC  using Adam. 
		\EndFor
		\State Set learning rate $l_r=d_r l_r$ if $i_\alpha/N_d$ is an integer.
		\EndFor
	\end{algorithmic}
\end{algorithm} 


\section{System Model with Imperfect CSI}\label{sec_impCH}
\begin{figure}[tb]  
	\centering
	\subfigure[Original system model with imperfect CSI. The angle $\theta_q$ is the 
	pre-processing angle defined in \eqref{eq_fb_alg_ahead}.]
	{\includegraphics[scale=.68, trim=0 0 0 0, clip]{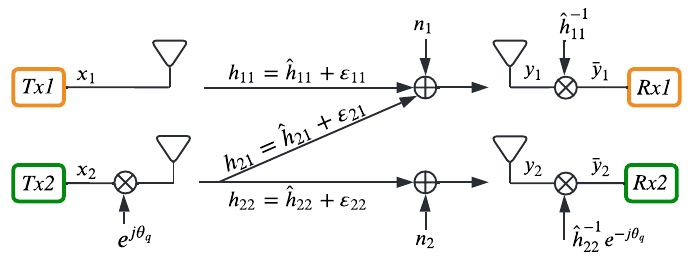}}
	\subfigure[Equivalent system model with imperfect CSI. The angle  
	$\theta_\delta$ is the 	quantization  	error defined in \eqref{eq_delt_theta}.] 
	{\includegraphics[scale=.68, trim=0 0 0 0, 
		clip]{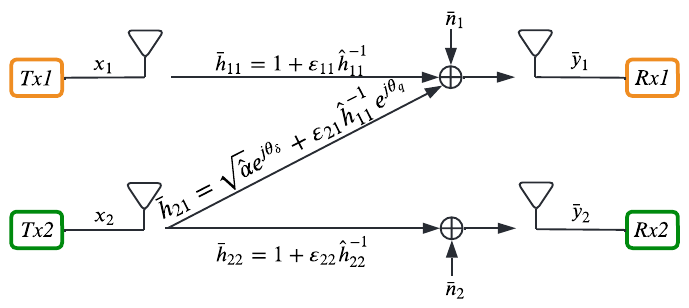}}
	\caption{System model of the ZIC with imperfect CSI.
	}
	\label{fig_sys_impCH}
\end{figure}

In this section, we consider the ZIC with \textit{imperfect CSI} and find its equivalent channel. The
CSI imperfectness comes from two sources: the error in the estimation of the CSI at each receiver and the error due to the quantization of the CSI before feeding it back to the transmitter. The  system model is depicted in Fig.~\ref{fig_sys_impCH}(a), in which both 
the estimation error and quantization error are considered. In general, the notation is similar to 
Section~\ref{sec_sysA}.
One main difference is that the estimation errors $\varepsilon_{ij}$ occur at the 
receivers when we estimate the 
channel coefficients $\hat{h}_{ij}$. The imperfectness of $\hat{h}_{ij}$ affects the decoding 
process. Besides, the 
quantization error occurs when the parameters need to be fed back to another node. For 
example, the \textit{Tx2} applies  
\begin{align}\label{eq_fb_alg_ahead}
	\theta_q \triangleq Q(\hat\theta_{11}-\hat\theta_{21}),
\end{align}
to cancel the phase of $h_{21}$ where $\theta_{11}$ and $\hat\theta_{21}$ are estimated by 
\textit{Rx1} and $Q(\cdot)$ is a quantization function. Ideally, $\theta_q$ is expected to be 
$\hat\theta_{11}-\hat\theta_{21}$, however, the quantization error exists which is defined as 
\begin{align}\label{eq_delt_theta}
\theta_\delta\triangleq\theta_q - (\hat\theta_{11}-\hat\theta_{21}),
\end{align}
The quantization error is included in $\bar{h}_{21}$ in Fig.~\ref{fig_sys_impCH}(b). The 
details of the system  model and the two types of errors are given as follows.

\subsubsection{Estimation Error}
The estimated channel gain is modeled as  in \cite{chen2005performance} 
\begin{align}\label{eq_impCH_org}
	\hat{h}_{ij}\triangleq	\hat{r}_{ij}e^{j\hat\theta_{ij}}={h}_{ij}-\varepsilon_{ij},
\end{align}
where 
\begin{align}\label{eq_impCH_dist}
	h_{ij}\sim \mathcal{CN}(\mu_h, \sigma_h^2)
\end{align}
is the actual channel  with 
mean $\mu_h$ and variance $\sigma_h^2$,  and
\begin{align}
	\varepsilon_{ij}\sim\mathcal{CN}(0,\sigma_E^2)\label{eq_estErrVar}
\end{align}
is the estimation error in which $\sigma_E^2$ 
is  the variance of the error, and $ij\in\{11,21,22\}$.
$\hat{h}_{11}$ and $\hat{h}_{21}$ 
are estimated by \textit{Rx1} while $\hat{h}_{22}$ is estimated by \textit{Rx2}. The 
estimated channel coefficients are determined by the actual channel and noise, hence, 
$\hat{h}_{ij}\sim\mathcal{CN}(\mu_h,\sigma_h^2+\sigma_E^2)$.
Once the actual 
channel and noise are determined, we have $h_{ij}=\hat{h}_{ij}+\varepsilon_{ij}$ as the channels 
in Fig.~\ref{fig_sys_impCH}(a).

\textit{Tx2} keeps the pre-processing based on a feedback angle, $\theta_q$.  \textit{Rx1} 
applies post-processing in Fig.~\ref{fig_sys2} based on the $\hat{h}_{ij}$. Then, 
$\bar{y}_i$ in \eqref{eq_ybar} becomes 
\begin{subequations}
	\begin{align}\label{eq_y_prime_1}
		\bar{y}_1=&\left(1+\frac{\varepsilon_{11}}{\hat{h}_{11}}\right)x_1
		+\frac{n_{1}}{\hat{h}_{11}}+e^{j\theta_q}\left(\frac{\hat{h}_{21}}{\hat{h}_{11}}
		+\frac{\varepsilon_{21}}{\hat{h}_{11}}\right)x_2,\\
		\bar{y}_2=&\left(1+\frac{\varepsilon_{22}}{\hat{h}_{22}}\right)x_2
		+\frac{n_{2}}{\hat{h}_{22}},
	\end{align}
\end{subequations}
which can be rewritten as 
\begin{subequations}
	\begin{align}\label{eq_y_prime_2}
		\bar{y}_1
		%
		=&\bar{h}_{11}x_1+\bar{h}_{21}x_2+\bar{n}_1,\\
		\bar{y}_2=&
		\bar{h}_{22}x_2+\bar{n}_2,
	\end{align}
\end{subequations}
by defining
\begin{subequations}
	\begin{align}
		&\bar{h}_{ii} \triangleq (1+\varepsilon_{ii} \hat{h}_{ii}^{-1}),\label{eq_impCh_dir}\\
		&\bar{h}_{21} \triangleq  (\hat{r}_{21}\hat{r}_{11}^{-1} e^{j\theta_\delta}+\varepsilon_{21} 
		\hat{h}_{11}^{-1}),\label{eq_impCh_cross}\\
		& \bar{n}_i\triangleq n_i\hat{h}_{ii}^{-1}\sim\mathcal{CN}(0,\sigma_N^2 
		\hat{r}_{ii}^{-2}).\label{eq_noisePw}
	\end{align}
\end{subequations}
{ In \eqref{eq_impCh_cross}, $\theta_\delta$  is the residual angle caused by the 
quantized feedback, which is defined in \eqref{eq_delt_theta} 
where ${\theta}_q$ is the feedback angle as \eqref{eq_fb_alg}. 

\subsubsection{Quantization Error}
Next, the transmitters require the knowledge of CSI to perform 
pre-processing and modification on constellation for enhanced performance. However, due 
to the limited feedback resources, the feedback information is quantized. This generates 
another source of imperfection.    The estimated and 
quantized parameters owned by 
each end are shown in Fig.~\ref{fig_sys_impCH_fb}. \textit{Rx1} has the estimated $r_{11}$, 
$r_{21}$, $\theta_{11}$, and $\theta_{21}$. \textit{Rx2} has the estimated $r_{22}$, 
and $\theta_{22}$. To make all the four nodes access to the channel knowledge, both 
\begin{subequations}
	\begin{align}
		&\alpha_q=Q(\hat\alpha)=Q({\hat{r}_{21}^2}\cdot{\hat{r}_{11}^{-2}}),\label{eq_fb_alp}\\
		&\color{black}\theta_q=Q(\hat\theta)=Q(\hat\theta_{11}-\hat\theta_{21})=\hat\theta_{11}
		-\hat\theta_{21}+\theta_\delta,\label{eq_fb_alg}
	\end{align} 
\end{subequations}
are sent to the transmitter and \textit{Rx2}, where  $Q(\cdot)$ is a uniform quantizer with 
accuracy $N_q$, $\hat\alpha$ is the estimation of the interference intensity,  
{$\hat{\theta}=\hat\theta_{11}-\hat\theta_{21}$}, and $\theta_\delta$ is the same  quantization 
error in \eqref{eq_delt_theta}. The quantizer $Q(\cdot)$ uniformly divides the 
region of the input variable  into $2^{N_q}$ 
segments. The region for $\alpha$ and angle are $[0,3]$ and $[-\pi, \pi]$, respectively. The 
middle value of the segment is the quantization result if the input value of 
$Q(\cdot)$ is within this segment. 
\begin{figure}[tbp] 
	\centering
	\includegraphics[scale=.62, trim=0 0 0 5 , 
	clip]{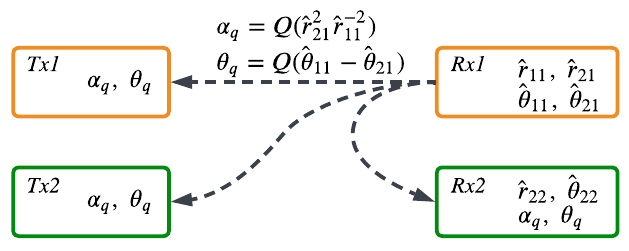}
	\caption{Feedback process for the  ZIC transmission. $\alpha_q$ and $\theta_q$ are the only feedback parameters.}
	\label{fig_sys_impCH_fb} 
\end{figure}

\subsubsection{Implementation of the imperfect ZIC}
To implement  the imperfect channel model, we randomly 
generate actual channels $h_{11}$ and $h_{22}$ and estimation errors $\varepsilon_{11}$, 
$\varepsilon_{22},$ and $\varepsilon_{21}$. Then,  $\hat{h}_{11}$ and $\hat{h}_{22}$ are 
determined by \eqref{eq_impCH_org}. We give interference gain $\alpha$ and 
then $h_{21}=\sqrt{\alpha}e^{i\theta_{21}}$, where $\theta_{21}$ is a random uniformly 
distributed 
angle on $[0,2\pi)$. {After  actual channels are generated, the receivers will have 
	the estimated CSI and can normalize the channel {as Fig.~\ref{fig_sys_impCH}(b)}. 
	The receivers 
	will then assume the channel 
	gains are unity and the \textit{Rx1} assumes the 
	interference gain is ${\hat\alpha}=|\hat{r}_{21}\hat{r}_{11}^{-1}|^2$. }

It is worth mentioning that,  $\bar{y}_{1}$ or $\bar{y}_{2}$ 
will go to infinity when the estimated channel gains ($r_{11}$ or $r_{22}$) are close to zero. In 
this case, the transmission is not reliable due to the wrong information 
obtained from the channel estimation no matter if the equalization in \eqref{eq_y_prime_2} is 
applied or not. The receivers will suffer from a mismatch between the received symbol 
and the constellation.  In the channel 
generation for the imperfect model, we keep the channel only if 
\begin{align}\label{eq_impCH_thres}
\max(\left|\frac{\varepsilon_{11}}{\hat{h}_{11}}\right|,  
\left|\frac{\varepsilon_{22}}{\hat{h}_{22}}\right|,  
\left|\frac{\varepsilon_{21}}{\hat{h}_{11}}\right|) <  T.
\end{align}
{where $T$ is a threshold. We use $T=1$ in this paper so that the estimation errors is 
	not dominating in $\bar{h}_{ii}$ in \eqref{eq_impCh_dir}.}

To summarize, for $i,j\in\{1,2\}$,
\begin{itemize}
	\item The actual channels coefficients are $h_{ij}$ in \eqref{eq_impCH_dist}.
	\item The estimated channels coefficients 
	are $\hat{h}_{ij}$, as in 
	\eqref{eq_impCH_org};
	\item After the equalization, the equivalent channel coefficients are $\bar{h}_{ij}$, as in 
	\eqref{eq_impCh_dir}-\eqref{eq_impCh_cross};
	\item After normalization, \textit{Rx1} and \textit{Rx2} will assume the direct channel 
	gains are one and the interference gain is $\hat{\alpha}$. 
	\item 	{\textit{Rx1} knows both the estimated and quantized 
	parameters. \textit{Rx1} sends feedback parameters ($\hat{\alpha}_q$ and $\hat{\theta}_q$)  
	to \textit{Tx1}, 	\textit{Tx2}, and 
	\textit{Rx2}.  
}
\end{itemize}

\begin{rem}
If $\sigma_E^2=0$ and $N_q \to \infty$ (i.e., the CSI is 
 perfect), 
the system in Fig.~\ref{fig_sys_impCH}(b) reduces to that of Fig.~\ref{fig_sys2}(b).
\end{rem}

\section{Deep Autoencoder for ZIC with Imperfect CSI}\label{sec_DAE_impCh}
The channel implementation should follow the imperfect ZIC model in 
Section~\ref{sec_impCH}.
The equivalent channels, $\bar{h}_{ij}$s, are set into the channel layers inside the DAE-ZIC. 
Different from the perfect channel case in \eqref{eq_DAE_ch1}-\eqref{eq_DAE_ch2} and 
Fig.~\ref{fig_2x2realCh}, the I/Q components  become
\begin{subequations}\label{eq_impCH_DAE}
	\begin{align}
	&\left[
	\begin{matrix}
	\bar{y}_{1}^{\textmd{I}}\\ \bar{y}_{1}^{\textmd{Q}}
	\end{matrix}
	\right]=\bar{\mathbf{H}}_{11}
	\left[
	\begin{matrix}
	x_{1}^{\textmd{I}}\\ x_{1}^{\textmd{Q}}
	\end{matrix}
	\right]+\bar{\mathbf{H}}_{21}
	\left[
	\begin{matrix}
	x_{2}^{\textmd{I}}\\ x_{2}^{\textmd{Q}}
	\end{matrix}
	\right]+\left[
	\begin{matrix}
	\bar{n}_{1}^{\textmd{I}}\\ \bar{n}_{1}^{\textmd{Q}}
	\end{matrix}
	\right],\\
	&\left[
	\begin{matrix}
	\bar{y}_{2}^{\textmd{I}}\\ \bar{y}_{2}^{\textmd{Q}}
	\end{matrix}
	\right]=\bar{\mathbf{H}}_{22}
	\left[
	\begin{matrix}
	x_{2}^{\textmd{I}}\\ x_{2}^{\textmd{Q}}
	\end{matrix}
	\right]+\left[
	\begin{matrix}
	\bar{n}_{2}^{\textmd{I}}\\ \bar{n}_{2}^{\textmd{Q}}
	\end{matrix}
	\right],
	\end{align}
\end{subequations}
in which $x_{i}^{\textmd{I}}$ and $x_{i}^{\textmd{Q}}$ are I/Q components of the transmitted 
signals, where 
$i\in\{1,2\}$ 
denotes the 
users; similarly, $\bar{y}_{i}^{\textmd{I}}$ and 
$\bar{y}_{i}^{\textmd{Q}}$ are I/Q components of the received signals; and,  
$\bar{n}_{i}^{\textmd{I}}$ and 
$\bar{n}_{i}^{\textmd{Q}}$ are the 
noises.  
$\bar{\mathbf{H}}_{ij}$ is  the real form of the complex-valued channel $\bar{h}_{ij}$ in 
\eqref{eq_impCh_dir}-\eqref{eq_impCh_cross}, 
\begin{align}\label{eq_net_impCH}
\bar{\mathbf{H}}_{ij}=\left[
\begin{matrix}
\bar{h}_{ij}^{\textmd{I}}& -\bar{h}_{ij}^{\textmd{Q}}\\
\bar{h}_{ij}^{\textmd{Q}}& \quad\bar{h}_{ij}^{\textmd{I}}
\end{matrix}
\right].
\end{align}
where $\bar{h}_{ij}=\bar{h}_{ij}^{\textmd{I}}+j\bar{h}_{ij}^{\textmd{Q}}$.  
Therefore, we add an FCNN layer  to implement each of the channels. 
The weight of the channel layer is $\bar{\mathbf{H}}_{ij}$ in \eqref{eq_net_impCH}. These channel layers have zero bias, no activation function, and are non-trainable. The weight will be 
replaced only if the channel is updated during the training and testing.  
\begin{figure}[tb] 
	\centering\hspace{-2.2mm}
	\includegraphics[scale=.48, trim=16 8 14 12,clip]{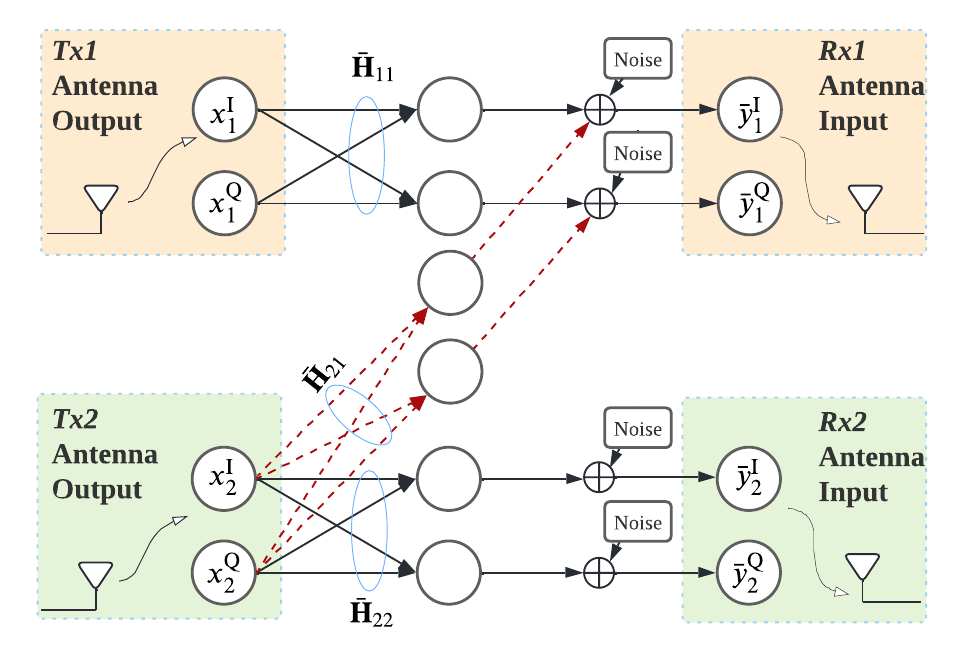}
	\caption{Implementation of the channels inside the proposed  DAE network. }
	\label{fig_DAE_ch} 
\end{figure}

{In contrast to the scenario with perfect CSI, illustrated in Fig.~\ref{fig_DAE_net}, the presence of imperfections in the CSI introduces distinct considerations to our approach. 
\begin{itemize}{
	\item Differences in parameters. In the case of imperfect CSI, the parameter $\alpha$ is not directly available for utilization.  The imperfectness at \textit{Rx1} side comes from the estimation of the interference gain $\hat{\alpha}$ and angle $\hat{\theta}$.  \textit{Rx1} transmits  $\alpha_q$ and $\theta_q$ (quantized values of $\hat{\alpha}$ and  $\hat{\theta}$)  to \textit{Tx1}, \textit{Tx2}, and \textit{Rx2}, as shown in Fig.~\ref{fig_sys_impCH_fb}. Thus, the information in three nodes suffer from both sources of  imperfectness, i.e., estimation and quantization errors.
	\item Differences in network input configurations. 
	 For \textit{Rx1}, the input comprises the non-quantized $\sqrt{\hat\alpha}$. Furthermore, \textit{Rx1} is uniquely positioned to incorporate the quantization error of the feedback angle, $\theta_\delta$, as an additional input. This is because \textit{Rx1}  oversees the quantization process, and thus is aware of the quantization discrepancy. As a result, $\theta_\delta$ serves as an input to the network at \textit{Rx1}.}
\end{itemize}
}
 The training process 
with imperfect channels is similar to Algorithm~\ref{alg_Train}. We still use 
Algorithm~\ref{alg_Train} but lines 10--13 of that algorithm will be replaced by  Algorithm~\ref{alg_impChGen}.

\begin{algorithm}[htbp]
	\caption{Channel Layer Preparation for DAE-ZIC}\label{alg_impChGen}
	\begin{algorithmic}[1]
		\State Inputs: $\mu_h$, $\sigma_h^2$, $\sigma_E^2$, $T$, and $N_q$.
		\While{1}
			\State Randomly generate $h_{11}$ and $h_{22}$ using \eqref{eq_impCH_dist}.
			\State Randomly generate $\varepsilon_{11}$, 
			$\varepsilon_{22}$, and $\varepsilon_{21}$ using \eqref{eq_estErrVar}.
			\If{\eqref{eq_impCH_thres} satisfied}
			\State Break.
			\EndIf
		\EndWhile
		\State {Uniformly generate  $\Delta\theta_q$ in $\frac{1}{2^{N_q}}[-\pi, \pi]$.
		\State Update $\theta_\delta$ using \eqref{eq_delt_theta}.
		\State Normalize the channels using \eqref{eq_impCh_dir} and \eqref{eq_impCh_cross}.}
		\State Set the channel layers  using \eqref{eq_impCH_DAE} and 
		according to Fig.~\ref{fig_DAE_ch}.
		\State Set the variance of the noise layer according to \eqref{eq_noisePw}.
		\State {Update  $\alpha_q$ and $\theta_q$ using \eqref{eq_fb_alp} and 
		\eqref{eq_fb_alg}. 
		\State Complete the feedback process according to Fig.~\ref{fig_sys_impCH_fb}.}
	\end{algorithmic}
\end{algorithm}

{
\section{Performance Analysis}\label{sec_simulation}
The performance is evaluated and compared for the three methods below:
\begin{itemize}
	\item \textit{DAE-ZIC}: The proposed method which designs new constellations based on the 
interference intensity.
	\item \textit{Baseline-1}: The transmitters directly use standard QAM. 
	\item \textit{Baseline-2}: \textit{Tx1} uses standard QAM, while \textit{Tx2} rotates 
	the standard QAM symbols based on the interference intensity \cite{knabe2010achievable, 
	ganesan2012two, deshpande2009constellation}.
\end{itemize}
First, we illustrate and analyze the constellations given by the proposed DAE-ZIC 
methods with perfect CSI. 
Then, the BER performance of the three methods is analyzed under perfect CSI assumption. 
Finally, we compare the performance of the three methods under imperfect channels, including 
imperfect 
estimation and quantization in the feedback.

\begin{figure*}[tbp]  
	\centering
	\subfigure[\textit{Baseline-1}, $\alpha=0.5$]{
		\includegraphics[scale=.38, trim=14 0 34 20, 
		clip]{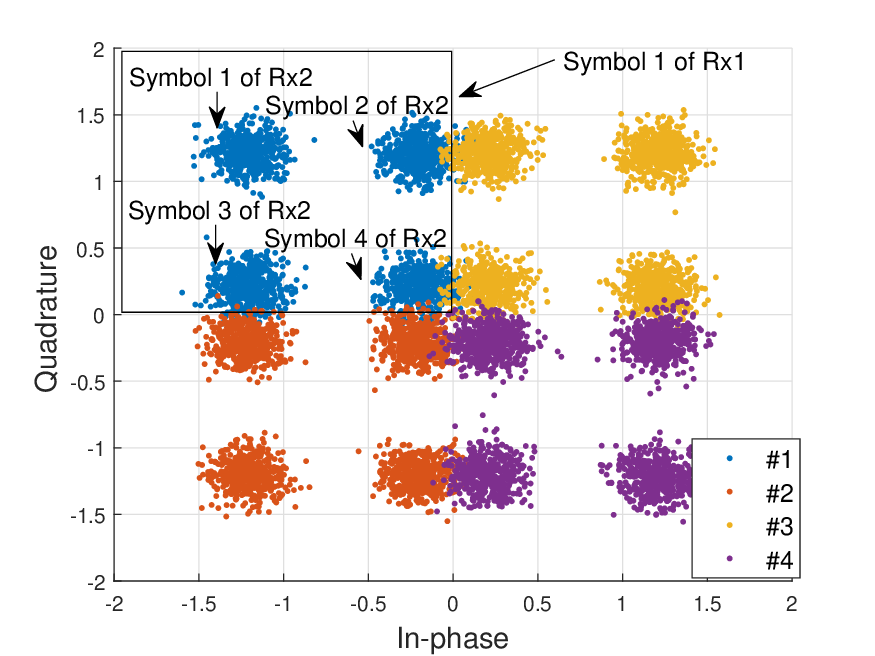}\label{fig_const_a1}}
	\subfigure[\textit{Baseline-2}, $\alpha=0.5$]{
		\includegraphics[scale=.38, trim=0 0 0 0, 
		clip]{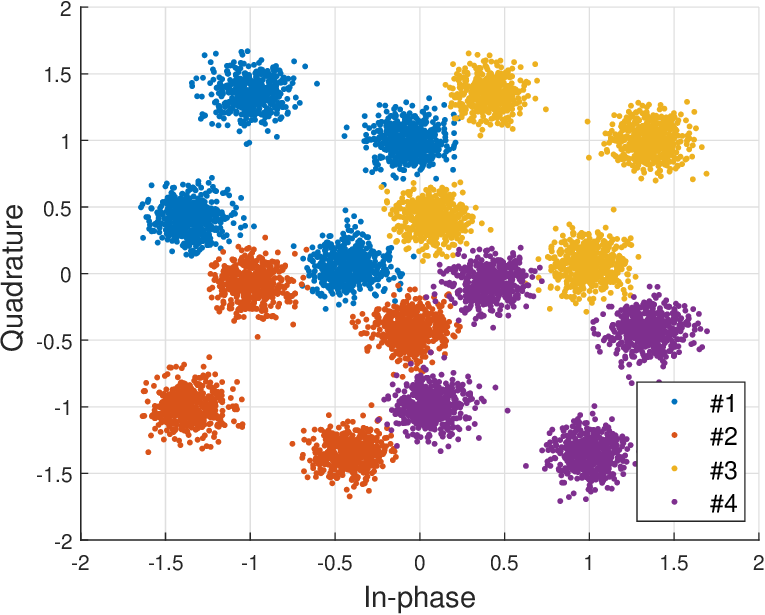}\label{fig_const_a2}}
	\subfigure[DAE-ZIC, $\alpha=0.5$]{
		\includegraphics[scale=.38, trim=0 0 0 0, 
		clip]{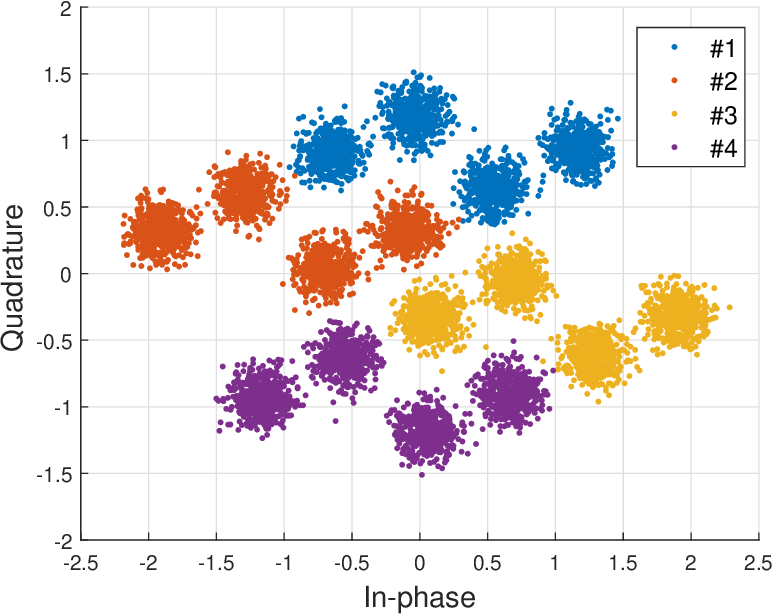}\label{fig_const_a3}}
	
	\subfigure[\textit{Baseline-1}, $\alpha=1$]{
		\includegraphics[scale=.38, trim=0 0 0 0, 
		clip]{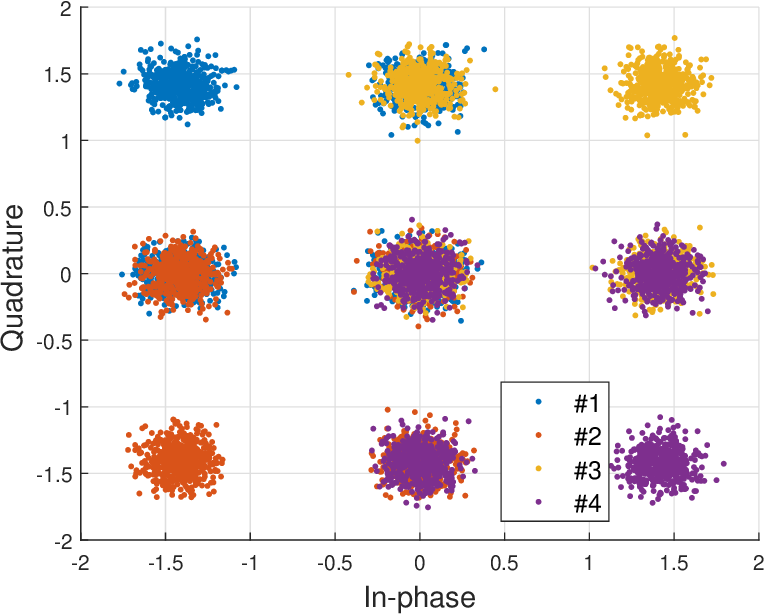}\label{fig_const_b1}}
	\subfigure[\textit{Baseline-2}, $\alpha=1$]{
		\includegraphics[scale=.38, trim=0 0 0 0, 
		clip]{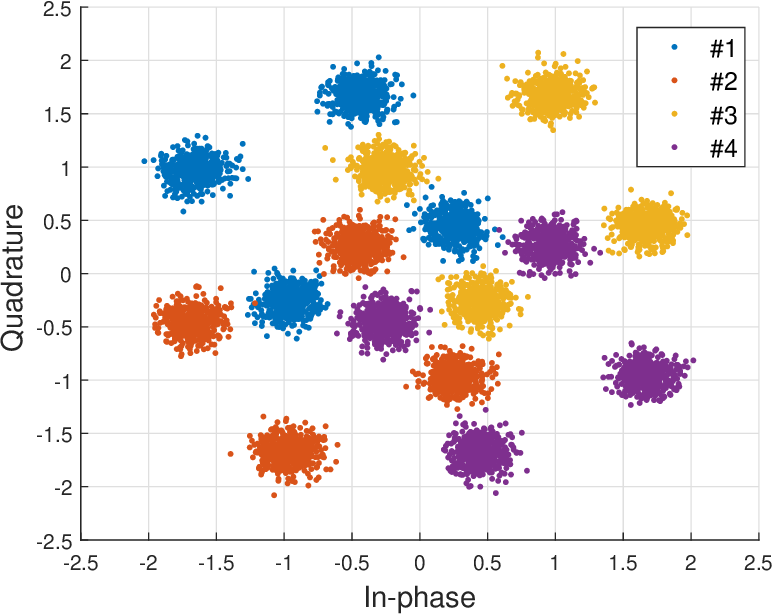}\label{fig_const_b2}}
	\subfigure[DAE-ZIC, $\alpha=1$]{
		\includegraphics[scale=.38, trim=0 0 0 0, 
		clip]{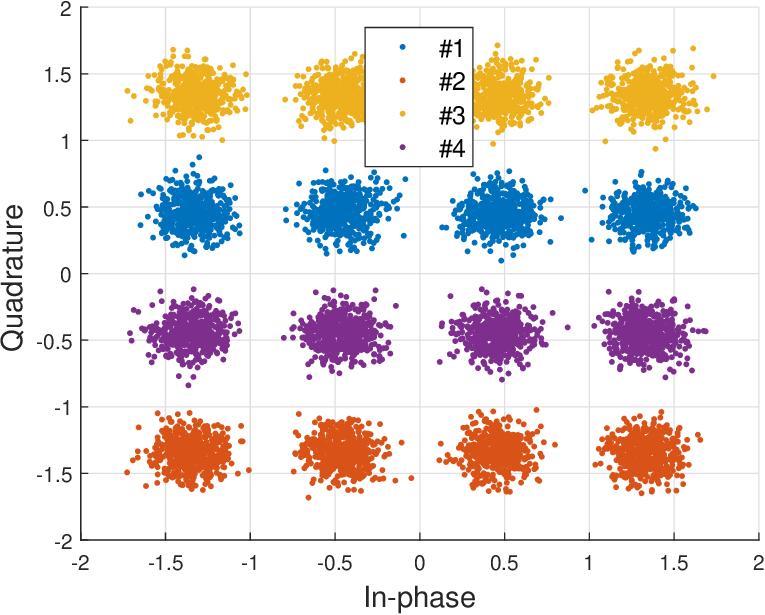}\label{fig_const_b3}}
	
	\subfigure[\textit{Baseline-1}, $\alpha=2$]{
		\includegraphics[scale=.38, trim=0 0 0 0, 
		clip]{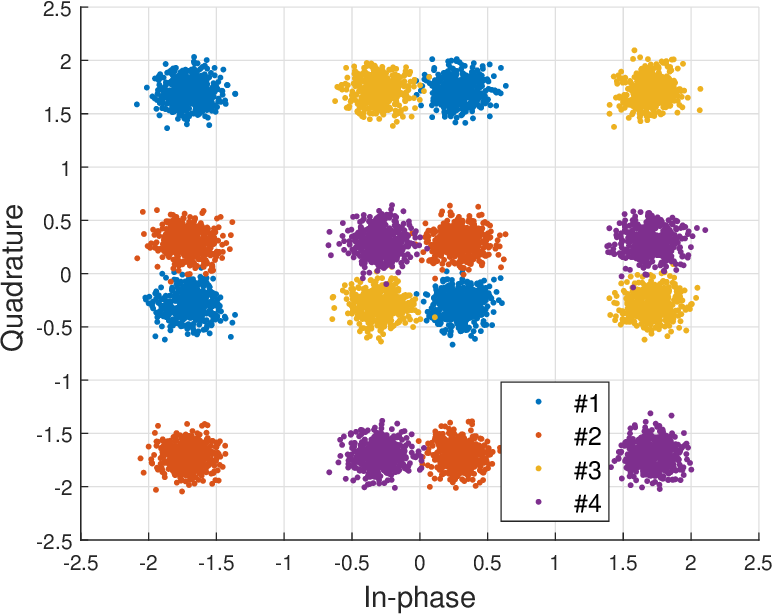}\label{fig_const_c1}}
	\subfigure[\textit{Baseline-2}, $\alpha=2$]{
		\includegraphics[scale=.38, trim=0 0 0 0, 
		clip]{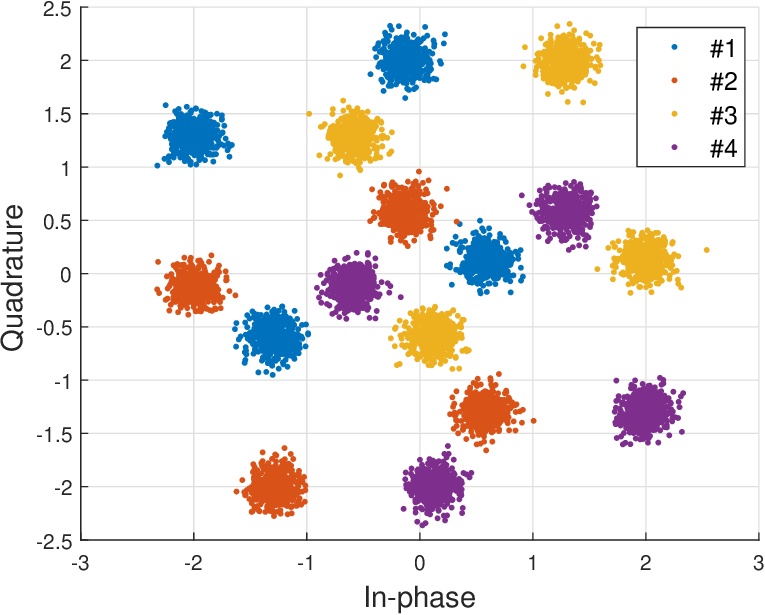}\label{fig_const_c2}}
	\subfigure[DAE-ZIC, $\alpha=2$]{
		\includegraphics[scale=.38, trim=0 0 0 0, 
		clip]{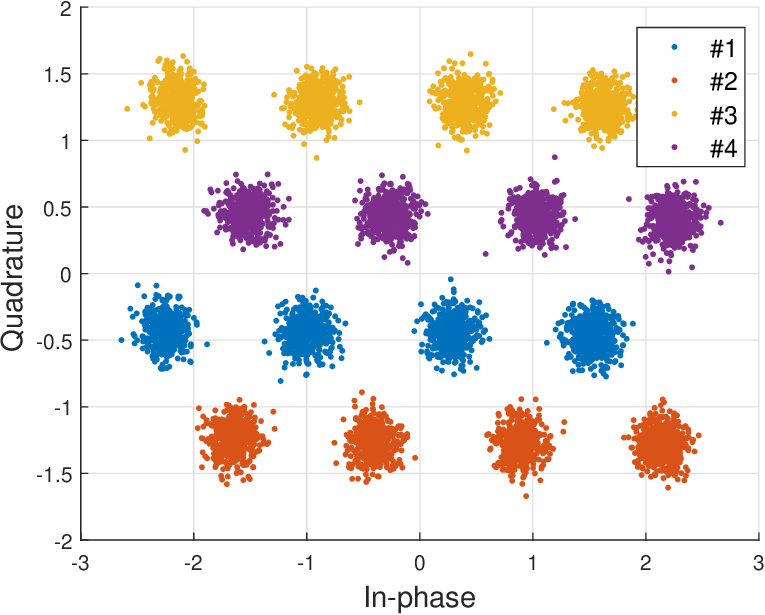}\label{fig_const_c3}}
	\caption{Constellation of  th DAE-ZIC and baselines at \textit{Rx1} for three 
		values of $\alpha$. The colors distinguish different symbols. Each of the four clusters of the 
		same color are caused by one of the four symbols from the interference with different noise.}
	\label{fig_const} 
\end{figure*}

\subsection{Network Ablation Study}
In this section, we conduct an ablation study to analyze the impact of network design and training parameters. Specifically, we investigate the effects of incorporating residual connections and selecting various training parameters. To this end, we design and perform six ablation experiments, which are detailed in Table~\ref{table:label}. The first experiment (Exp.~1) examines the presence or absence of shortcuts in the proposed method. 
Through Exp.~2 to Exp.~5, we investigate the impact of feeding or not feeding the training parameter $\sqrt{\alpha}$ into different sub-networks at the transmitters and receives. Exp.~6 explores the importance of Sub-network~2 which is a new design to
perform power allocation.

\begin{table}[h]
	\centering
	\caption{Ablation settings and their performance for different values of $\alpha$.} 
	\label{table:label}
	\scriptsize
	\begin{tabular}{l@{\hskip 0.08in}c@{\hskip 0.08in}c@{\hskip 0.08in}c@{\hskip 0.08in}c@{\hskip 0.08in}c@{\hskip 0.08in}c@{\hskip 0.08in}c@{\hskip 0.08in}c@{\hskip 0.08in}|}
		\hline
		\textbf{Settings} & \textbf{Proposed} & \textbf{Exp.~1} & \textbf{Exp.~2} & \textbf{Exp.~3} & \textbf{Exp.~4} & \textbf{Exp.~5}&\textbf{Exp.~6} \\ \hline
		\textbf{Use shortcuts} & \checkmark & - & \checkmark & \checkmark & \checkmark & \checkmark & \checkmark\\ \hline
		\textbf{$\sqrt{\alpha}$ to sub-net~1} & \checkmark & \checkmark & - & \checkmark & - & \checkmark & \checkmark\\ \hline
		\textbf{$\sqrt{\alpha}$ to sub-net~2} & \checkmark & \checkmark & \checkmark & - & - & \checkmark & -\\ \hline
		\textbf{$\sqrt{\alpha}$ to the Rx} & \checkmark & \checkmark & \checkmark & \checkmark & \checkmark & - & \checkmark\\ \hline
		\textbf{Use sub-net~2} & \checkmark & \checkmark & \checkmark & \checkmark & \checkmark & \checkmark & - \\ \hline
		\hline
		\textbf{Performance} & \textbf{Proposed} & \textbf{Exp.~1} & \textbf{Exp.~2} & \textbf{Exp.~3} & \textbf{Exp.~4} & \textbf{Exp.~5} &  \textbf{Exp.~6} \\ \hline
	\boldmath$\alpha=0.5$ & 0.0252 &   0.0424 &   0.0324 &   0.0318   & 0.0340   & 0.0513  &  0.0934 \\ \hline
	\boldmath$\alpha=1$ & 0.0236 &   0.1207 &   0.0294  &  0.0288 &   0.0281  &  0.0488   & 0.0391 \\ \hline
		\boldmath$\alpha=1.5$ & 0.0212 &   0.1288 &   0.0253  &  0.0237  &  0.0308  &  0.0463  &  0.0328 \\ \hline
	\end{tabular}
\end{table}


	\subsubsection{The impact of residual connections} The incorporation of residual connections into the proposed DAE has a significant impact on its performance, especially for bit inputs. The addition of these shortcuts enables the network to increase its learning capacity and improve its performance without the need for additional parameters or a wider network. Moreover, these skip connections help preserve gradients, allowing the network to learn representations at varying depths. As can be seen  in the first two columns of Table~\ref{table:label}, excluding residual connections significantly increases the BER,  indicating that residual connections play a crucial role in the overall performance of the proposed DAE-ZIC.

\subsubsection{The impact of $\sqrt{\alpha}$ at the Txs} We develop three experiments (Exp.~2, Exp.~3, and Exp.~4) in which  $\sqrt{\alpha}$ is fed into  either sub-networks or non and compare their BER with the proposed network in which $\sqrt{\alpha}$ is fed into  bot sub-networks. While there is not big performance difference between Exp.~2, Exp.~3, and Exp.~4,  for all $\alpha$ settings, there is a significant improvement  when $\sqrt{\alpha}$ is fed into  both sub-networks (proposed method). 
Sub-network~1 primarily focuses on constellation design, whereas Sub-network~2 regulates the power of I/Q dimension. Since  $\sqrt{\alpha}$ influences both the  constellation layout and the power distribution, providing $\sqrt{\alpha}$ to each sub-network is essential for optimizing the overall performance of the system.  

\subsubsection{The impact of $\sqrt{\alpha}$ at the Rx}
Exp.~5 examines the performance when $\sqrt{\alpha}$ is not fed to the decoder. This experiment emphasizes the importance of incorporating interference information at the decoder side, as otherwise the BER roughly doubles (compared to the first column). 

\subsubsection{The impact of Sub-network~2} Exp.~6 demonstrates the importance of Sub-network~2, which is a new  design  to ensure different powers at I/Q dimensions. Our simulations reveal that the DAE-designed constellations downgraded to PSK in the absence of Sub-network~2.   This explains why neglecting Sub-network~2 yields results similar to the \textit{Baseline-2} which uses rotated QPSK.
On the contrary, as shown by the simulations, the inclusion of Sub-network~2 allows the proposed structure to design other shapes such as pulse-amplitude keying (PAM) at each transmitter. The combination of these PAMs ultimately yields a QAM-like constellation at the receiver affected by interference. These findings further emphasize the importance of incorporating Sub-network2, as it enables effective power control across the I/Q dimensions, crucial for ensuring the proper functioning of the constellations.

\subsection{Scenario I: Perfect CSI}
\begin{figure*}[htbp] 
	\centering
	\subfigure[${M}_1={M}_2=4$, $\alpha=0.5$]{
		\includegraphics[scale=.44, trim=0 0 0 2, 
		clip]{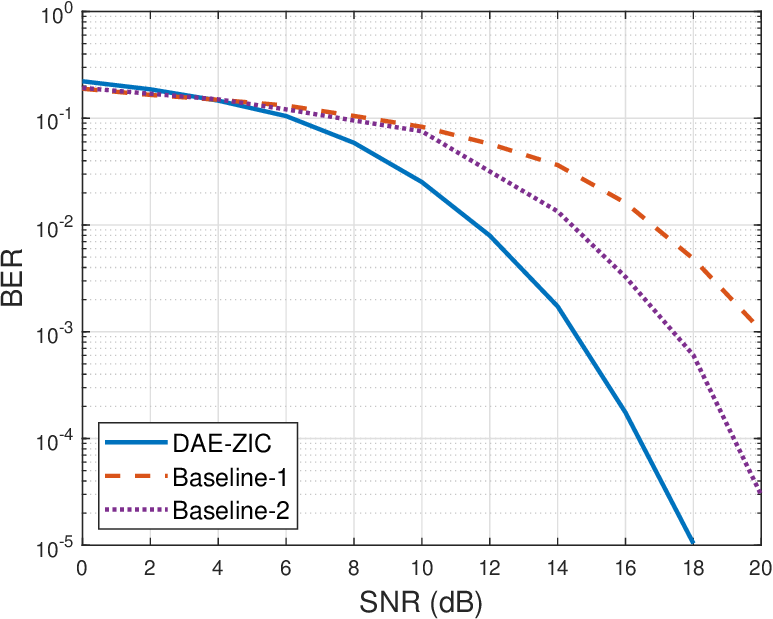}\label{fig_cmpSNR_a2}}\hspace{-2.1mm}
	\subfigure[${M}_1={M}_2=4$, $\alpha=1$]{
		\includegraphics[scale=.44, trim=0 0 0 0, 
		clip]{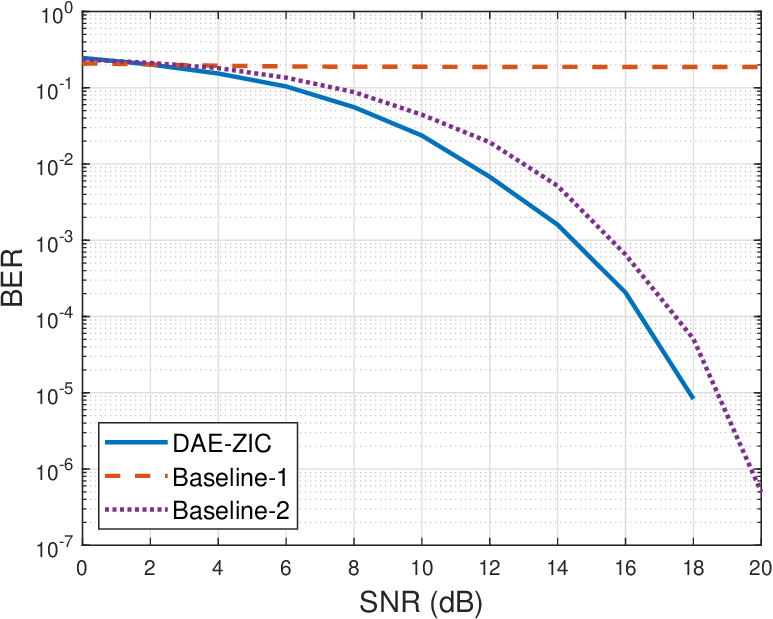}\label{fig_cmpSNR_a3}}\hspace{-2.1mm}
	\subfigure[${M}_1={M}_2=4$, $\alpha=1.5$]{
		\includegraphics[scale=.44, trim=0 0 0 0, 
		clip]{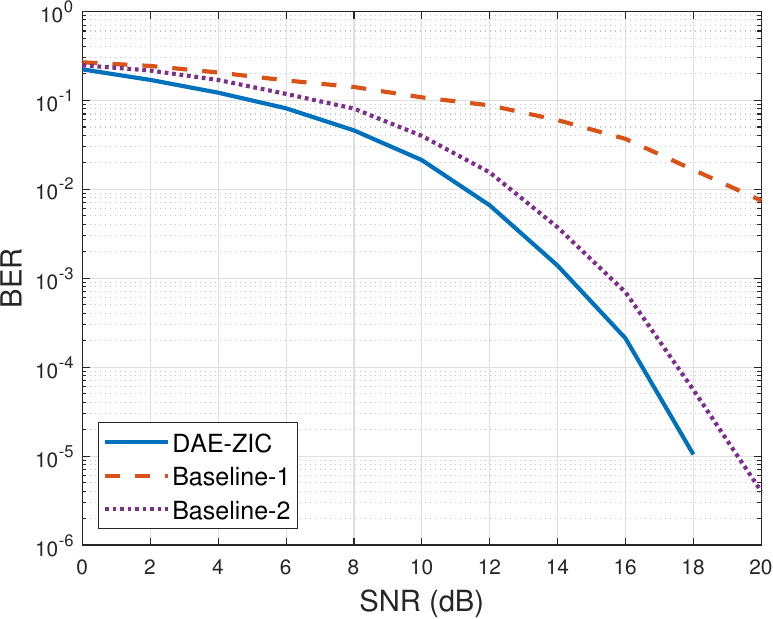}\label{fig_cmpSNR_a4}}\hspace{-2.1mm}
	
	\subfigure[${M}_1={M}_2=8$, $\alpha=0.5$]{
		\includegraphics[scale=.44, trim=0 0 0 0, 
		clip]{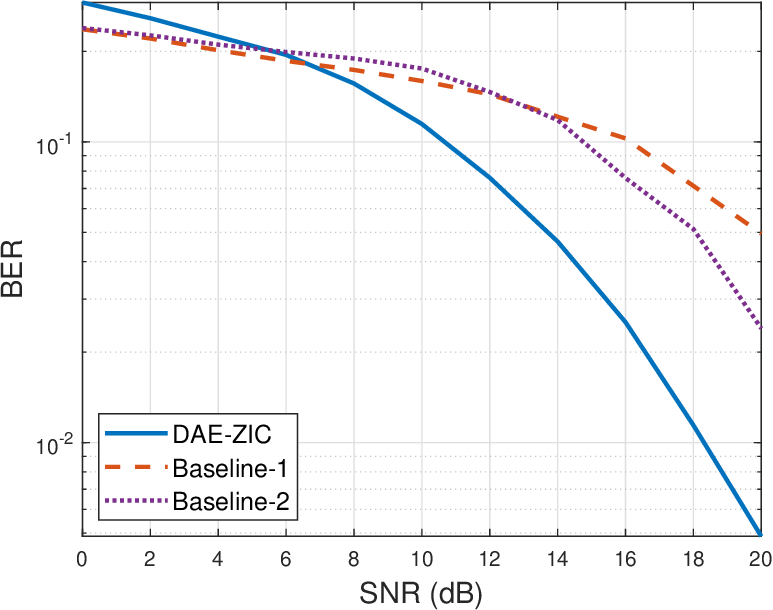}\label{fig_cmpSNR_a8}}\hspace{-2.1mm}
	\subfigure[${M}_1={M}_2=8$, $\alpha=1$]{ 
		\includegraphics[scale=.44, trim=0 0 0 0, 
		clip]{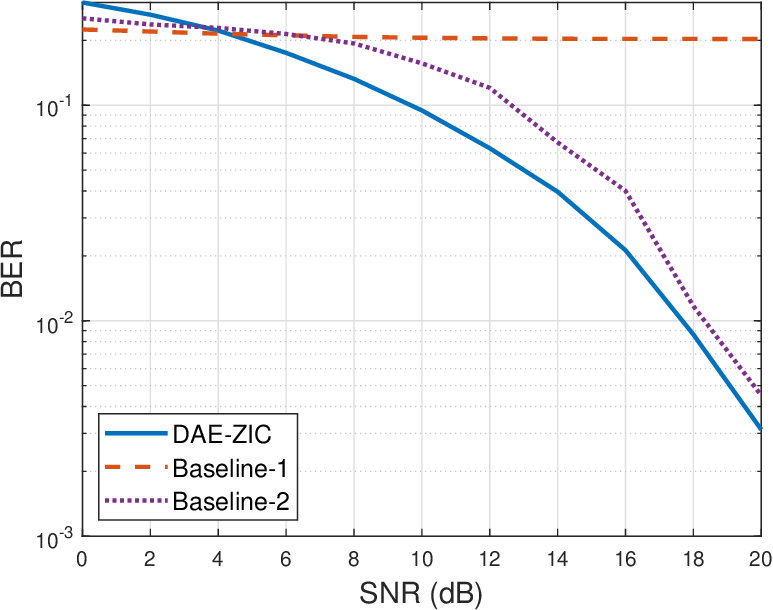}\label{fig_cmpSNR_a9}}\hspace{-2.1mm}
	\subfigure[${M}_1={M}_2=8$, $\alpha=1.5$]{  
		\includegraphics[scale=.44, trim=0 0 0 0, 
		clip]{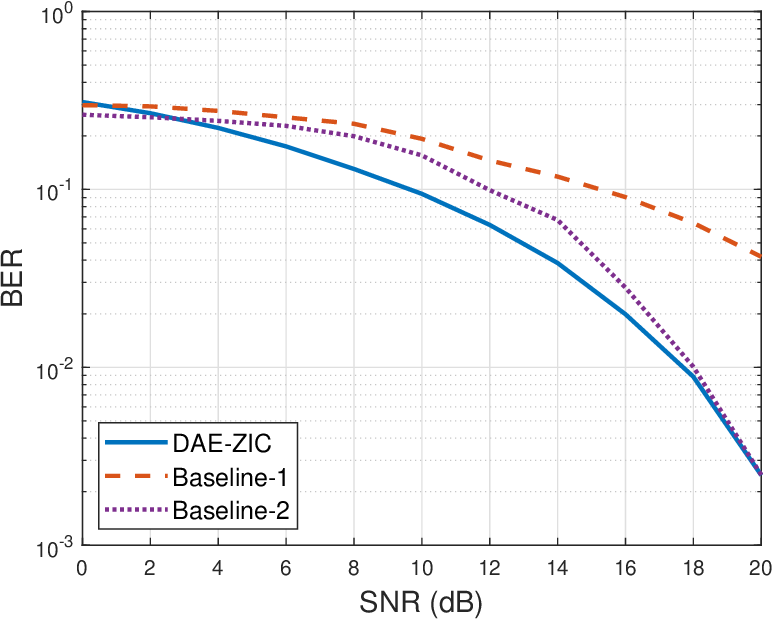}\label{fig_cmpSNR_a10}}\hspace{-2.1mm}
	\caption{The maximum (worst case) BER performance among the two users  of 
		DAE-ZIC 
		versus SNR. The BER is averaged over different interference gains  which is  
		uniformly 
		distributed in the given interval.}
	\label{fig_cmpSNR_a} 
\end{figure*}
\begin{figure}[htbp] 
	\centering
	\subfigure[${M}_1={M}_2=4$]{
		\includegraphics[scale=.45, trim=0 0 0 0, 
		clip]{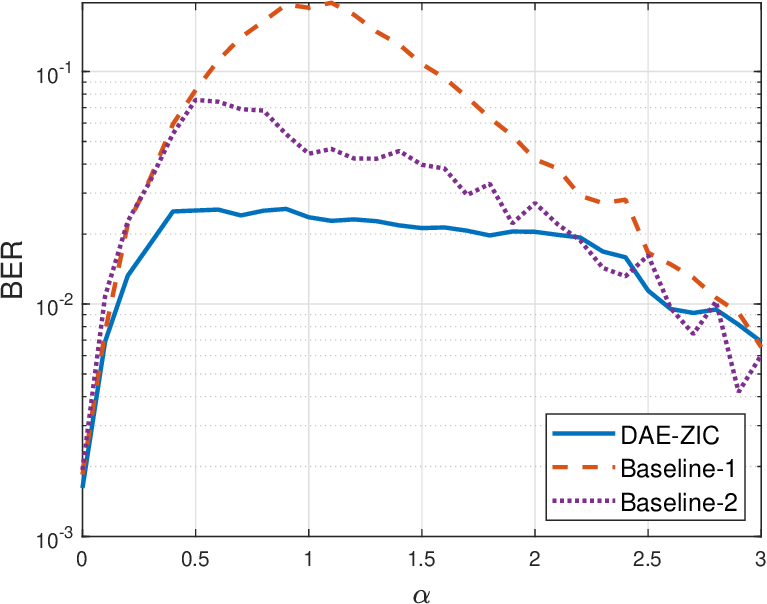}\label{fig_cmpSlice_a1}}\hspace{-2.1mm}
	\subfigure[${M}_1={M}_2=8$]{
		\includegraphics[scale=.45, trim=0 0 0 0, 
		clip]{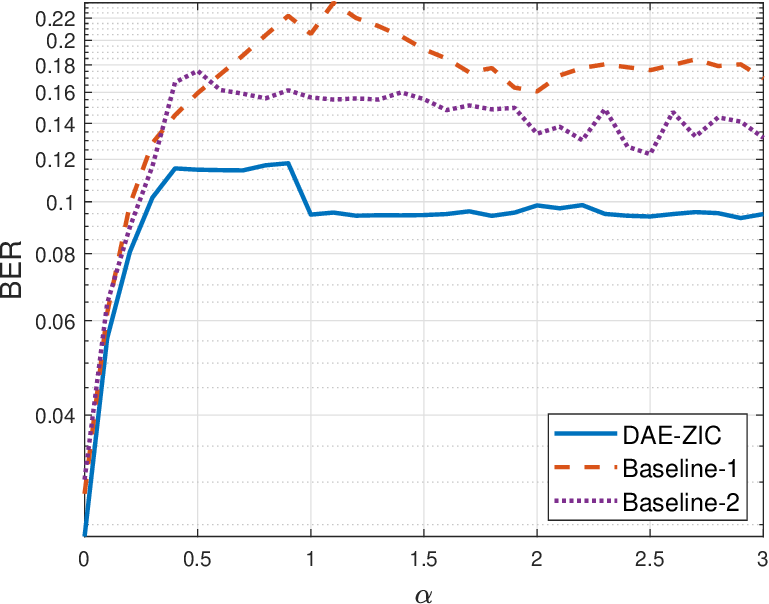}\label{fig_cmpSlice_a4}}\hspace{-2.1mm}
	\caption{The maximum (worst case) BER performance 
		versus  interference gains. The SNR is fixed as 10dB.}
	\label{fig_cmpSlice_a} 
\end{figure}
\begin{figure*}[tbp] 
	\centering
	\subfigure[${M}_1={M}_2=4$, $\alpha=0.5$]{
		\includegraphics[scale=.44, trim=0 0 0 18, 
		clip]{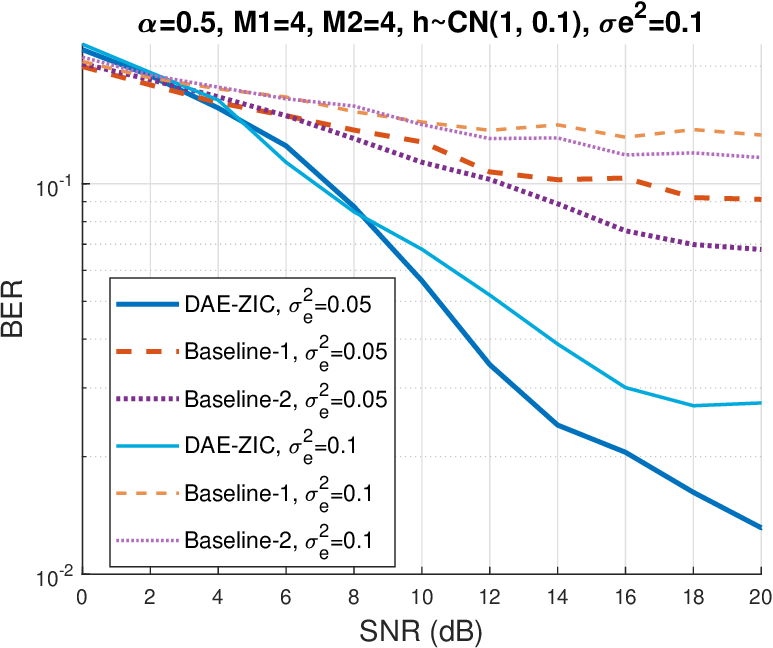}\label{fig_cmpSNR_a0.5}}\hspace{-2.1mm}
	\subfigure[${M}_1={M}_2=4$, $\alpha=1$]{
		\includegraphics[scale=.44, trim=0 0 0 18, 
		clip]{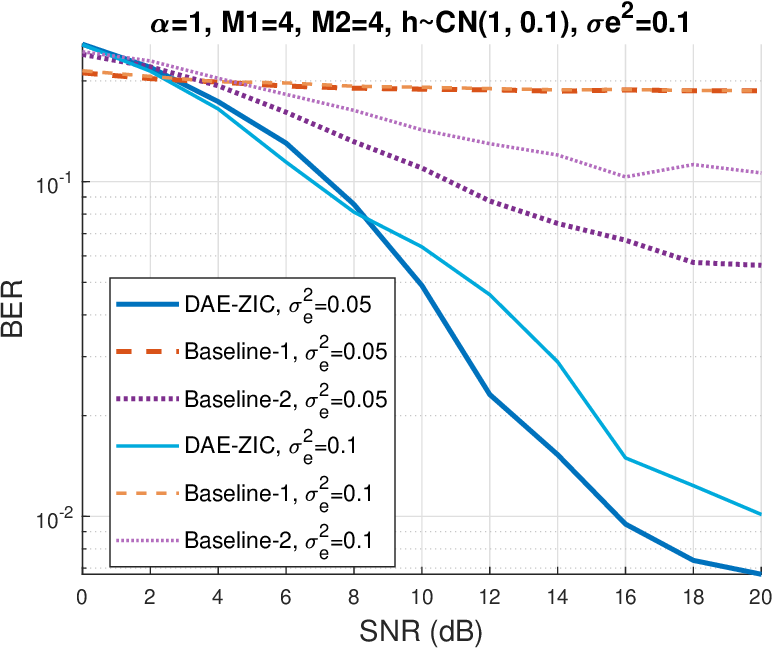}\label{fig_cmpSNR_a1}}\hspace{-2.1mm}
	\subfigure[${M}_1={M}_2=4$, $\alpha=1.5$]{
		\includegraphics[scale=.44, trim=14 0 0 22, 
		clip]{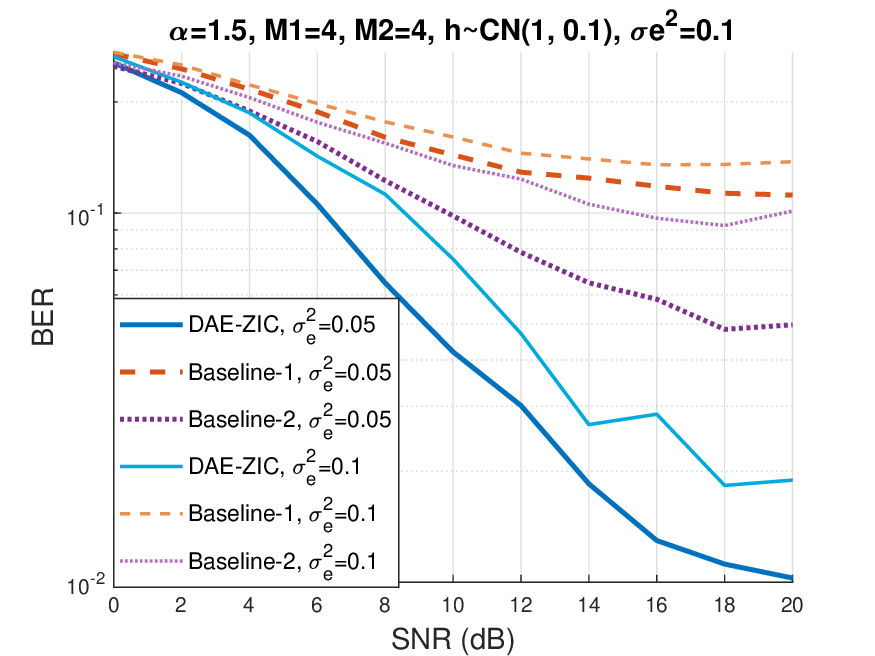}\label{fig_cmpSNR_a1.5}}\hspace{-2.1mm}
	\caption{The maximum (worst) BER performance among the two users  versus SNR when
		estimation error is considered.}
	\label{fig_cmpSNR_1h0.1} 
\end{figure*}
\subsubsection{Constellation Analysis}
The received constellations at \textit{Rx1} for all three methods (the two baselines and the proposed DAE-ZIC) are shown 
in Fig.~\ref{fig_const}. {In this simulation,} we set $N_s=2$ so that each user has $2^{N_s}=4$  
information symbols. The channels are perfectly known, the transmit power is unity, and the SNR is 8dB. 

Each sub-figure of Fig.~\ref{fig_const} contains four four-symbol clusters differentiated 
by 
different colors. Each cluster refers to a  symbol transmitted to \textit{Rx1}. Within each  
cluster, there are four symbols, which are due to the symbols of \textit{Rx2}. 
For example, the blue colors denote symbol $1$ of \textit{Rx1}, each contaminated by one of the symbols  of \textit{Rx2} (interference) and AWGN noise. It can be seen 
that the location and distribution of symbols  are different in each method.  
The constellations of \textit{Baseline-1} (left column) 
are very crowded and even overlapped when 
 $\alpha=1$ in Fig~\ref{fig_const_b1}. {This is because \textit{Tx2} strongly interferes with the transmission of \textit{Tx1} to \textit{Rx1}} 
 by directly applying 4-QAM.   \textit{Baseline-2} (middle column)  
 rotates the constellation of \textit{Tx2}, which enlarges the space between symbols and thus 
 makes the 
decoding easier. The proposed DAE-ZIC (right-column) creates the most separable constellations. 
It can better 
make use of the I/Q plane in constellation design based on the interference intensity. The two 
cooperating DAEs can intelligently choose and adjust various scaled constellation types to avoid  
constellation overlapping. 
When $\alpha=0.5$, the DAE-ZIC designs  parallelogram-shape constellations
compared with the 
square-shape constellations in the baselines. For 
$\alpha=1$, both the constellation of both \textit{Tx1}  and \textit{Tx2} morph to PAM. The two PAM constellations are 
perpendicular to each other hence the overlapping is eliminated. When 
$\alpha=2$, \textit{Tx1} uses a parallelogram-shape QAM and \textit{Tx2} uses a PAM. 
By adapting their constellations to the interference intensity, 
the two DAEs cooperate to avoid constellation overlap as much as possible. This is the main 
reason that the DAE-ZIC outperforms the baselines. 

In addition, compared with \cite{wu2020deep}, where the symbols exhibit nearly equal power in one stream, our design allows the I/Q domains to have different power levels by incorporates sub-network 2.  With this, the transmitter can generate QAM-like constellations, further mitigating the impact of interference. It is also worth noting that imperfect CSI is not considered in \cite{wu2020deep}, we will show in the next section.  

\subsubsection{BER Performance of the DAE-ZIC}
To evaluate the effectiveness of the DAE-ZIC, we compare the BER of the three 
methods over SNR in $[0,20]$dB and $0\leq\alpha\leq3$. 
For the DAE-ZIC, we divide interval $\alpha\in[0,3]$ in to six sub-intervals, i.e, $[0,0.5)$, 
$[0.5,1)$, 
$\hdots$, $[2.5,3]$. For each sub-interval of $\alpha$, we train a  
DAE-ZIC through Algorithm~\ref{alg_Train}.

The BER versus the SNR  is shown in Fig.~\ref{fig_cmpSNR_a}. In each sub-figure, 
	$\alpha$  is a fixed value.  In general, the DAE-ZIC outperforms 
	the two baseline models, especially in moderate and strong interference regimes.
With $N_s=3$, performance of DAE-ZIC drops at 0dB. The reason is we have trained the network 
at $\textmd{SNR}=10$dB but have tested it for a range of SNRs from 0 to 20dB. A potential way 
to improve is to train the DAE-ZIC with a variety of SNRs.

The  BER performance versus $\alpha$ at  $\textmd{SNR}=10$dB is 
shown in Fig.~\ref{fig_cmpSlice_a}. In Fig.~\ref{fig_cmpSlice_a}(a) and 
Fig.~\ref{fig_cmpSlice_a}(b), we set $N_s=2$ and $N_s=3$, i.e.,  
${M}_1={M}_2=4$ and ${M}_1={M}_2=8$. 
When  interference is very weak, i.e., $\alpha\in[0,0.25]$, the three methods have similar BERs. 
The proposed DAE-ZIC noticeably reduces the BER in weak, moderate, and strong interference 
cases, where $\alpha\in[0.5, 2]$. From Fig.~\ref{fig_cmpSlice_a}(a),  using DAE-ZIC  over $\alpha\in[0,3]$,
BER is reduced {75.7\% and 44.29\%} with respect to \textit{Baseline-1} and \textit{Baseline-2}, respectively. When 
$\alpha\in[0.5, 2]$, the improvement becomes {80.3\% and 51.5\%}. When the interference is very strong, e.g., $\alpha>2.5$, \textit{Baseline-2} slightly outperforms the DAE-ZIC. The reason could be 
that 4-QAM with rotation may achieve  optimal performance \cite{knabe2010achievable}.
For $N_s=3$ in Fig.~\ref{fig_cmpSlice_a}(b),  {44.4\% and 31.5\%} BER reduction is reached 
by the DAE-ZIC 
over $\alpha\in[0, 3]$. 
For $\alpha\in[0.5,2]$, DAE-ZIC outperforms the other two methods with {45.0\%  and 
33.4\%}. 
\textit{Baseline-1}  performs poorly for $\alpha\in[0.5,2]$, because the two 
added QAM constellations may get crossed and overlap. 
Thus, \textit{Rx1} cannot decode its message. Such a phenomenon is alleviated in 
\textit{Baseline-2} which simply rotates one QAM constellation and the added constellations still 
have a reasonable symbol distance. The normalization layer allows 
the DAE-ZIC to design constellations without any regular-shape restrictions. Thus, the minimum 
distance at receivers can be enlarged which results in a lower BER.

\subsection{Scenario II: Imperfect CSI}
The influences of the imperfect CSI on the BER performance are shown as 
follows. Two types of imperfection are analyzed  independently: CSI estimation error and 
quantization error.  {For imperfect 
	CSI, both baselines use the system model in 
	Fig.~\ref{fig_sys_impCH}(b). When 
	decoding, \textit{Rx1} will take $\theta_\delta$ into decoding.}  The performance of the two 
baselines and the 
proposed DAE-ZIC are evaluated for  $\alpha\in[0,3]$ and SNR in $[0,20]$dB. In the simulation, 
the actual direct channel  $h_{ii}\sim\mathcal{CN}(1,0.1)$ where $i\in\{1,2\}$. The estimation error 
$\varepsilon_{ij}$s are in $\mathcal{CN}(0,\sigma_E^2)$. Then the estimated channels are 
obtained from \eqref{eq_impCH_org}. 
To be fair to the users, 
we use the maximum (worst) BER between them as the measurement. Each result is 
averaged over 500 random channels.

\begin{figure}[tbp] 
	\centering
	\subfigure[${M}_1={M}_2=4$]{
		\includegraphics[scale=.45, trim=0 0 0 19, 
		clip]{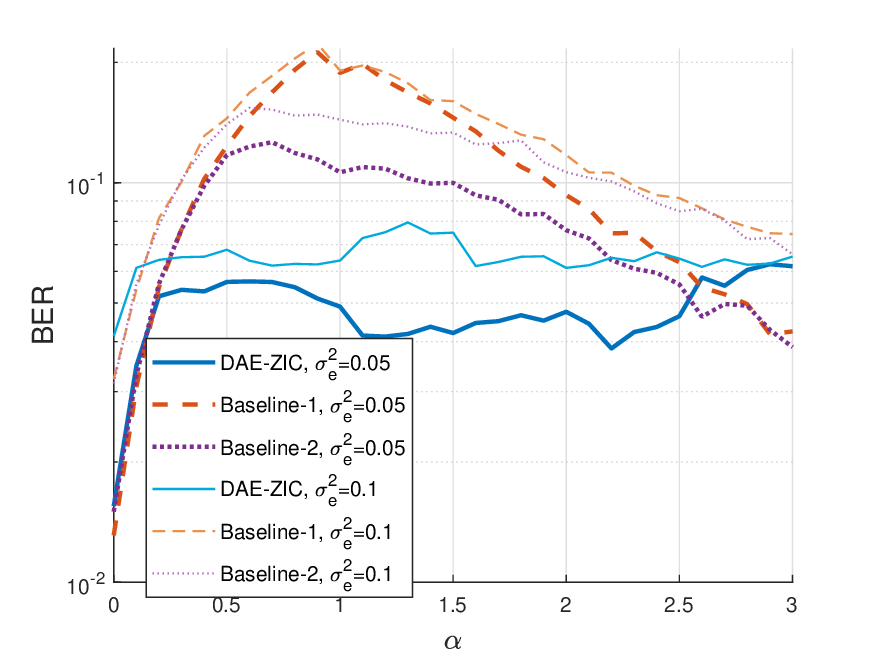}\label{fig_cmpSlice_1h0.1_M4}}\hspace{-2.1mm}
	\subfigure[${M}_1={M}_2=8$]{
		\includegraphics[scale=.45, trim=0 0 0 19, 
		clip]{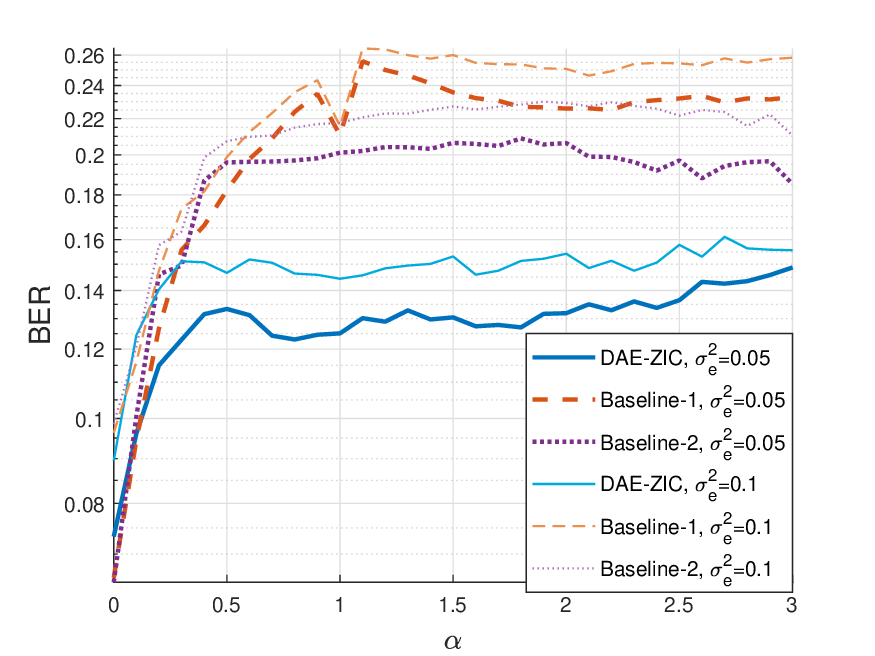}\label{fig_cmpSlice_1h0.1_M8}}\hspace{-2.1mm}
	\caption{The maximum (worst) BER performance of the two users under the estimation 
		error. The SNR is fixed as 10dB.}
	\label{fig_cmpSlice_1h0.1} 
\end{figure}

\begin{figure*}[tbp] 
	\centering 
	\subfigure[${M}_1={M}_2=4$, $\alpha=0.5$]{
		\includegraphics[scale=.44, trim=0 0 0 0, 
		clip]{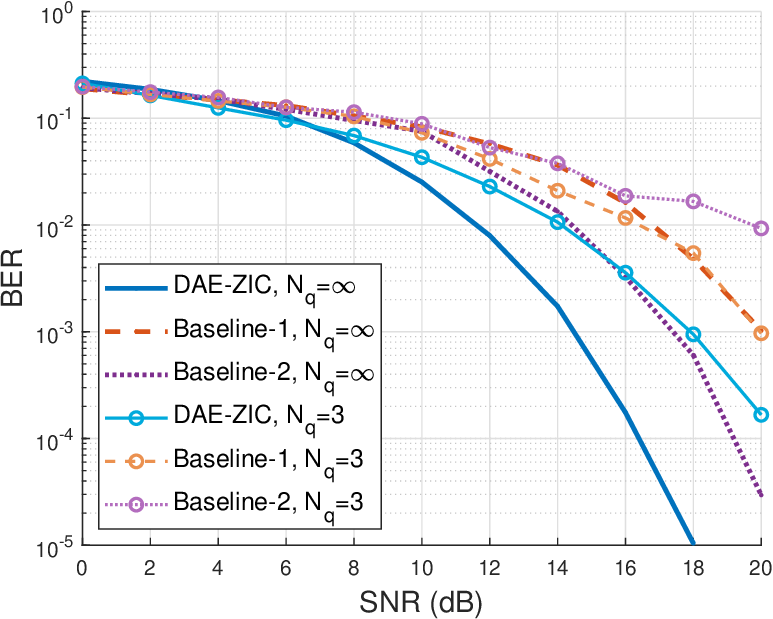}\label{fig_cmpSNRqt_a0.5}}\hspace{-2.1mm}
	\subfigure[${M}_1={M}_2=4$, $\alpha=1$]{
		\includegraphics[scale=.44, trim=0 0 0 0, 
		clip]{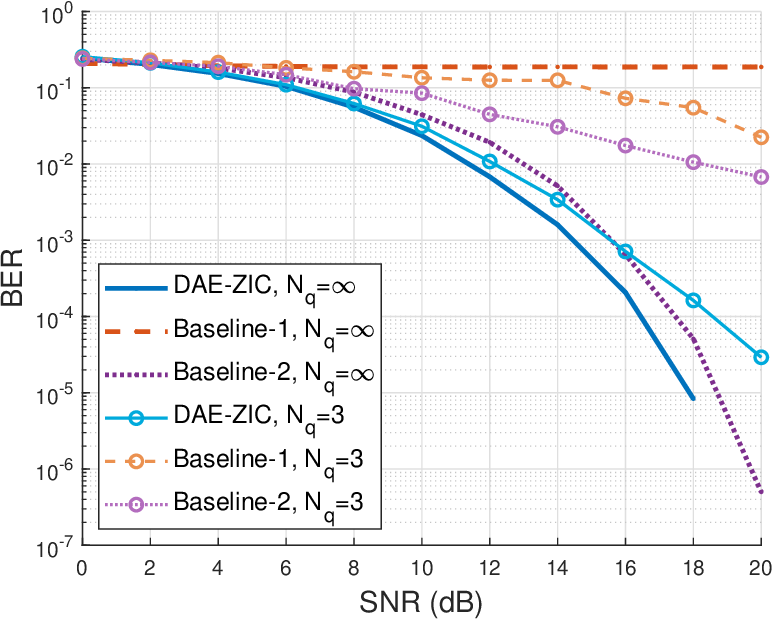}\label{fig_cmpSNRqt_a1}}\hspace{-2.1mm}
	\subfigure[${M}_1={M}_2=4$, $\alpha=1.5$]{
		\includegraphics[scale=.44, trim=0 0 0 0, 
		clip]{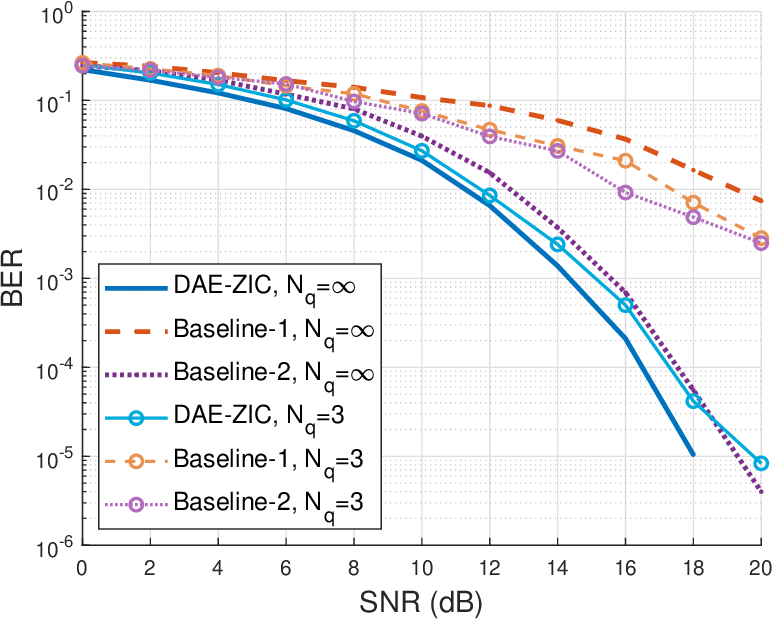}\label{fig_cmpSNRqt_a1.5}}\hspace{-2.1mm}
	\caption{The maximum (worst) BER performance of the two users  versus SNR. The 
		quantization is considered while free from estimation error.}
	\label{fig_cmpSNRqt_1h0.1} 
\end{figure*}
\subsubsection{CSI with Estimation Error}
The BER versus interference intensity $\alpha$ is shown in 
Fig.~\ref{fig_cmpSlice_1h0.1}. Two levels of the estimation error are examined. Specifically, 
$\sigma_E^2\in\{0.05, 0.1\}$. The proposed DAE-ZIC outperforms the baselines almost for any 
$\alpha$.  The percentage of the  BER reduction
is shown in Table~\ref{tab_impch_BER}. With a fixed $\alpha$, the BER 
versus SNR is shown in Fig.~\ref{fig_cmpSNR_1h0.1} for weak, moderate, and strong 
interference. Compared to the baseline methods, the DAE-ZIC has a remarkable improvement in BER performance in all  interference and estimation error levels. As mentioned earlier, the 
main difference between the proposed DAE-ZIC and the baselines is that the network is able to 
re-design the constellation based on the interference intensity, and that is how the DAE-ZIC reduces the 
BER.

\begin{table}[htbp]
	\caption{BER   Reductions of  DAE-ZIC   Compared to the Baseline Methods 
	(in Percentage and $\alpha\in[0,3]$). }\label{tab_impch_BER}
\centering
\begin{tabular}{l|l|cc|cc}
	\hline
	\multicolumn{2}{l|}{Compared to}  & \multicolumn{2}{c}{\textit{Baseline-1}} & 
	\multicolumn{2}{c}{\textit{Baseline-2}} \\ 
	\hline
	\multicolumn{2}{l|}{$N_s$}                    & 2          & 3          & 2          & 3          \\ \hline
	\multirow{3}{*}{$\sigma_E^2$}
	&$0$   &  75.77\%   &  44.29\%     & 44.43\% & 31.50\%   \\
	&$0.05$ &  55.40\%  &   38.97\%    & 39.12\% &  31.43\%   \\
	&$0.1$  &   48.83\%  &  35.81\%    & 41.41\% &    29.24\% \\ \hline        
\end{tabular}
\end{table}

\subsubsection{CSI with Feedback Quantization}
{When channel estimation is perfect, the BER performance  with feedback quantization (with ${N_q=3}$) and without quantization (${N_q=\infty}$)  is shown in Fig.~\ref{fig_cmpSNRqt_1h0.1}.  
The DAE-ZIC outperforms the baselines with and without feedback quantization.
In addition, the quantization error increases the BER of all methods. However, the performance degradation of the DAE-ZIC is much less than the other two methods. Especially, for $\alpha=1$ and $\alpha=1.5$), where the interference is strong, the DAE-ZIC outperform other methods by more than two orders of magnitude when  $N_q=3$  and $\rm SNR=20$dB.}  
Interestingly, \textit{Baseline-1} performs  better for $N_q=3$ compared to $N_q=\infty$. The reason is that quantization of the angles may  introduce rotation angle on \textit{Tx2}, acting like \textit{Baseline-2} unintentionally. This then  reduces the constellation overlapping and thus a better BER is reached.
 
{With high interference intensities, i.e,  $\alpha=1$ and $\alpha=1.5$, the BER of the DAE-ZIC with quantization ($N_q=3$) is only slightly degraded compared to the un-quantized case ($N_q=\infty$).  
This is because, while \textit{Tx}s receive the quantized CSI, \textit{Rx1} knows CSI before and after quantization. Then, it  can to some extent correct the imperfectness in transmitters.  Therefore, there is no dramatic degradation for the DAE-ZICs.}  
{On the other hand,  \textit{Baseline-2} is more sensitive to the quantization error and thus a big gap of the BER happens between $N_q=\infty$ and $N_q=3$.} 
The reason is that  \textit{Baseline-2} only rotates the constellation in  
\textit{Tx2}, which highly depends on the phase shifted by the channels.}  

\section{Conclusion}\label{sec_con}
A novel DAE architecture for interference mitigation in the two-user ZIC with perfect and imperfect CSI has been proposed.  The DAE-ZIC  minimizes the  BER by 
jointly designing transmit and receive DAEs and optimizing them together. In this architecture, the average power constraint 
is realized by designing a normalization layer.  This enables the proposed DAE-ZIC  to design more efficient 
symbols to achieve lower BERs.  BER simulations verify the effectiveness of the proposed 
structure.  We have compared the proposed DAE-ZIC with two baseline models, and the DAE-ZIC outperforms both. With quantized CSI, the gain obtained by the DAE-ZIC compared to the best conventional method is remarkable and can be as large as two orders of magnitude at $\rm SNR = 20 dB$. 
Getting back to the questions asked in the introduction, we conclude that autoencoder is  a viable solution for interference channels and it outperforms the conventional methods by designing new, nonuniform constellations which make the symbols separable at the interfered  
receiver side.

\begin{IEEEbiography}
	[{\includegraphics[width=1in,height=1.25in,clip,keepaspectratio]{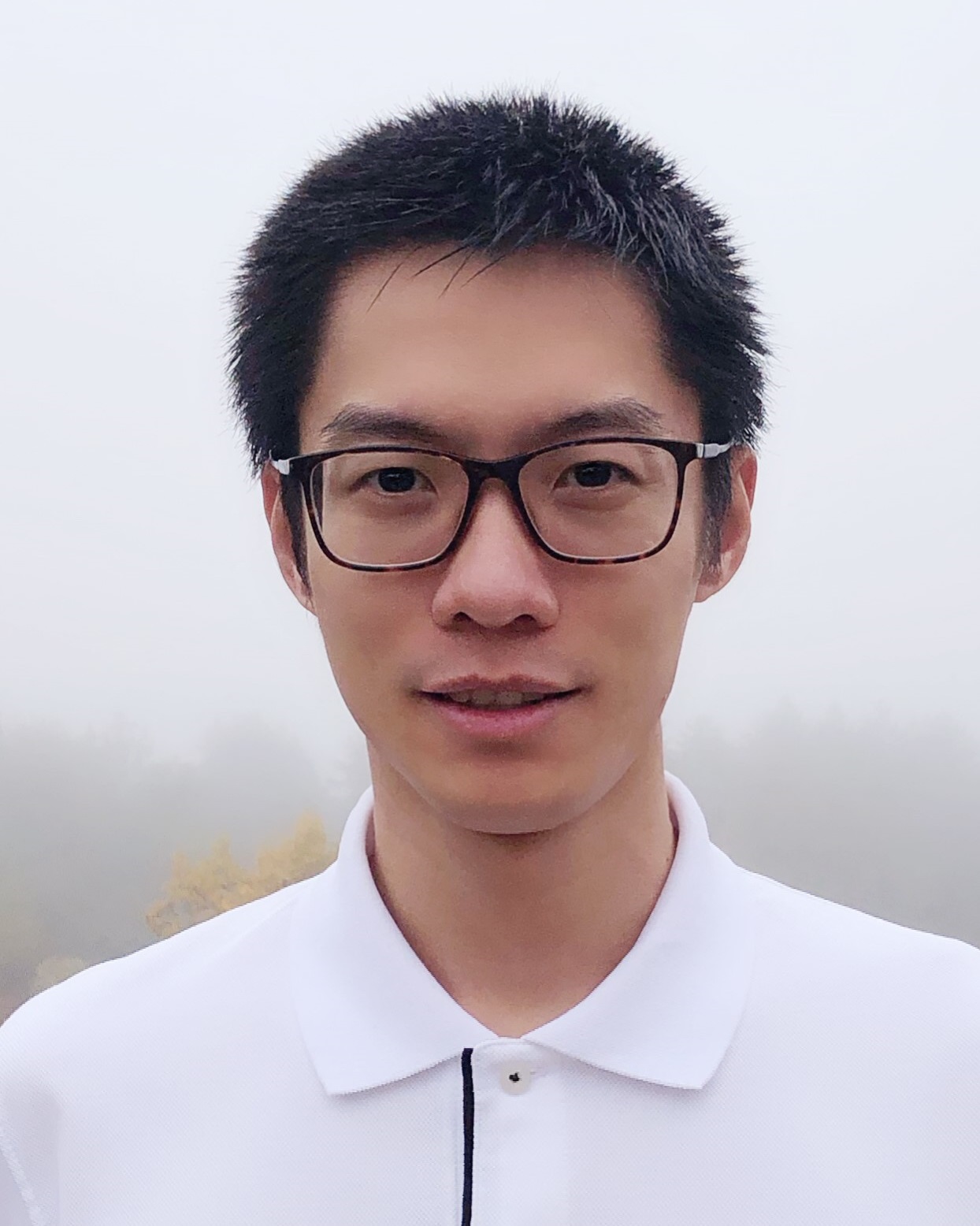}}]{Xinliang
		Zhang}
	(S'19) received the B.Eng. and M.Eng. degrees in Electrical Engineering 
	from Xidian University, Xi'an, China, in 2015 and 2018, respectively. He obtained the Ph.D. degree 
	from the Department of Electrical and Computer Engineering, Villanova 
	University in 2022. His research interests lie in physical layer 
	security, machine learning for wireless communications, MIMO networks, 
	and signal processing.

\end{IEEEbiography}

\begin{IEEEbiography}
	[{\includegraphics[width=1in,height=1.25in,clip,keepaspectratio]{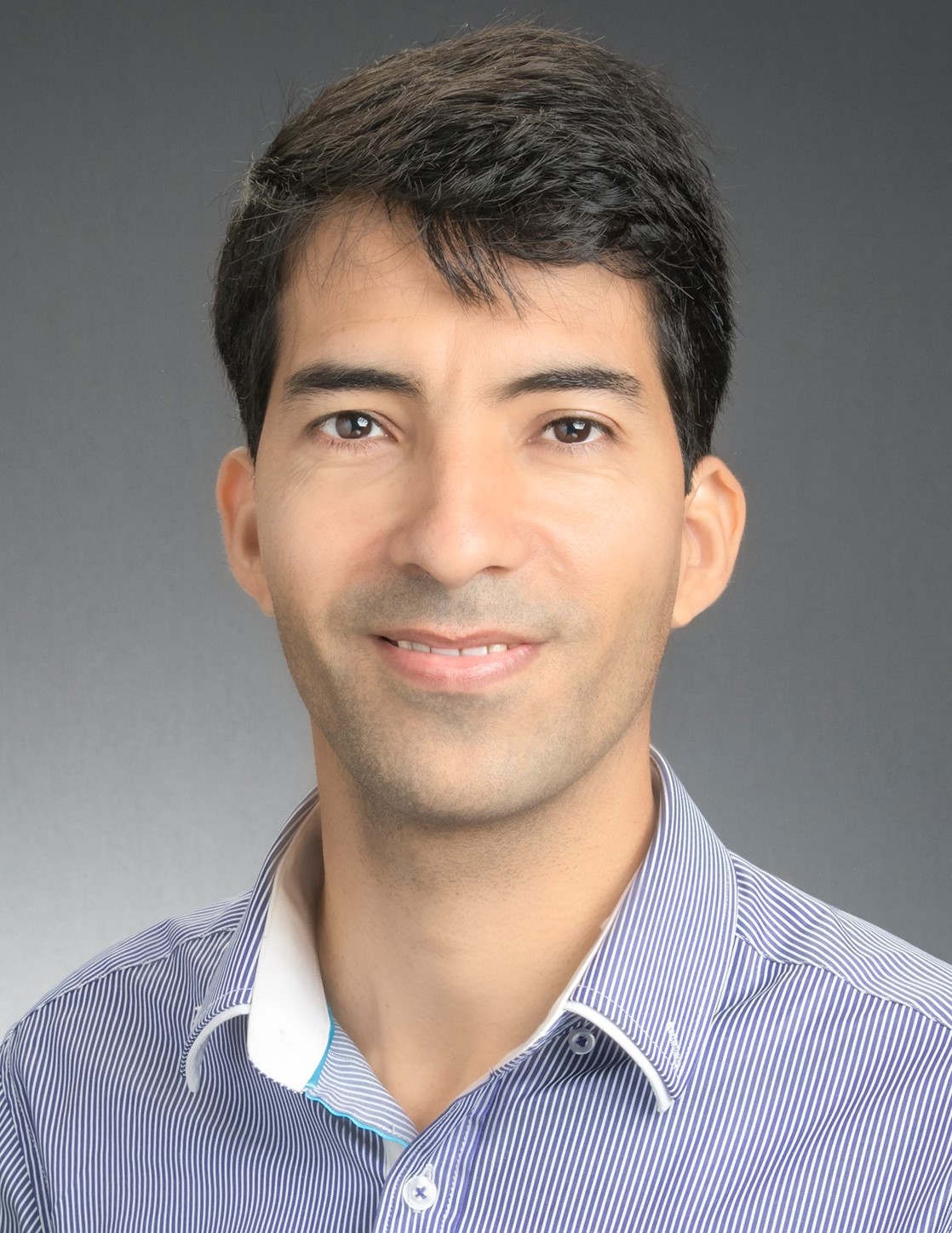}}]{Mojtaba Vaezi} (S’09–M’14–SM’18) received the B.Sc. and M.Sc. degrees from Amirkabir University of Technology (Tehran Polytechnic) and the Ph.D. degree from McGill University, all in Electrical Engineering. From 2015 to 2018, he was with Princeton University as a Postdoctoral Research Fellow and Associate Research Scholar. He is currently an Associate Professor of ECE at Villanova University. Before joining Princeton, he was a researcher at Ericsson Research in Montreal, Canada. His research interests include the broad areas of signal processing and machine learning for wireless communications with an emphasis on physical layer security and sixth-generation (6G) radio access technologies. Among his publications in these areas is the book \textit{Multiple Access Techniques for 5G Wireless Networks and Beyond} (Springer, 2019). 
	
	Dr. Vaezi has held editorial positions at various IEEE journals, currently serving as an Editor for \textsc{IEEE Transactions on Communications}  along with his role as a Senior Editor for \textsc{IEEE Communications Letters} and  Senior Area Editor for \textsc{IEEE Signal Processing Letters}.  He is a recipient of several academic, leadership, and research awards, including McGill Engineering Doctoral Award, IEEE Larry K. Wilson Regional Student Activities Award in 2013, the Natural Sciences and Engineering Research Council of Canada (NSERC) Postdoctoral Fellowship in 2014, Ministry of Science and ICT of Korea’s best paper award in 2017, IEEE Communications Letters Exemplary Editor Award in 2018,  the 2020 IEEE Communications Society Fred W. Ellersick Prize,  the 2021 IEEE Philadelphia
	Section Delaware Valley Engineer of the Year Award, and the National Science Foundation (NSF) CAREER Award in 2023. 
\end{IEEEbiography}

\end{document}